\documentclass{aa}  

\usepackage{graphicx}
\usepackage[varg]{txfonts}
\usepackage[colorlinks=true,     linkcolor=blue, citecolor=blue, filecolor=blue, urlcolor=blue]{hyperref}

\begin{document}

\title{The $M_\bullet$--$\sigma_e$ relation for local type~1 AGNs and quasars}

\author{J. Molina\inst{1,2}
            \and L. C. Ho\inst{3,4}
            \and K. K. Knudsen\inst{2}
          }

\institute{Instituto de F\'isica y Astronom\'ia, Universidad de Valpara\'iso, Avda. Gran Breta\~na 1111, Valpara\'iso, Chile
                                \and Department of Space, Earth and Environment, Chalmers University of Technology, SE-412 96 Gothenburg, Sweden
              \and Kavli Institute for Astronomy and Astrophysics, Peking University, Beijing 100871, People's Republic of China
              \and Department of Astronomy, School of Physics, Peking University, Beijing 100871, People's Republic of China\\
              \email{juan.molinato@uv.cl}
          }

\date{Received XXXXXX XX, 2023; accepted XXXXX XX, 2023}

\abstract{We analyzed Multi Unit Spectroscopic Explorer observations of 42 local $z\lesssim0.1$ type~1 active galactic nucleus (AGN) host galaxies taken from the Palomar-Green quasar sample and the close AGN reference survey. Our goal was to study the relation between the black hole mass ($M_\bullet$) and bulge stellar velocity dispersion ($\sigma_e$) for type~1 active galaxies. The sample spans black hole masses of $10^{6.0}-10^{9.2}\,M_\odot$, bolometric luminosities of $10^{42.9}-10^{46.0}\,$erg\,s$^{-1}$, and Eddington ratios of $0.006-1.2$. We avoided AGN emission by extracting the spectra over annular apertures. We modeled the calcium triplet stellar features and measured stellar velocity dispersions of $\sigma_* = 60-230$\,km\,s$^{-1}$ for the host galaxies. We find stellar velocity dispersion values in agreement with previous measurements for local ($z\lesssim0.1$) AGN host galaxies, but slightly lower compared with those reported for nearby X-ray-selected type~2 quasars. Using a novel annular aperture correction recipe to estimate  $\sigma_e$ from $\sigma_*$ that considers the bulge morphology and observation beam-smearing, we estimate flux-weighted $\sigma_e = 60-250$\,km\,s$^{-1}$. If we consider the bulge type when estimating $M_\bullet$, we find no statistical difference between the distributions of AGN hosts and the inactive galaxies on the $M_\bullet$--$\sigma_e$ plane for $M_\bullet \lesssim 10^8\,M_\odot$. Conversely, if we do not consider the bulge type when computing $M_\bullet$, we find that both distributions disagree. We find no correlation between the degree of offset from the $M_\bullet$--$\sigma_e$ relation and Eddington ratio for $M_\bullet \lesssim 10^8\,M_\odot$. The current statistics preclude firm conclusions from being drawn for the high-mass range. We argue these observations support notions that a significant fraction of the local type~1 AGNs and quasars have undermassive black holes compared with their host galaxy bulge properties.}

\maketitle

\section{Introduction}
The discovery of the relations between host galaxy bulge properties and the mass of central supermassive black holes (BHs; \citealt{Kormendy1995,Magorrian1998,Ferrarese2000,Gebhardt2000}) has been pivotal in extragalactic astronomy during the last two decades. Nowadays, it is widely accepted that all galaxies with a massive bulge host a supermassive BH, with the BHs being a key element in regulating galaxy formation and evolution, so that the growth of both components is intimately related \citep[see also \citealt{Greene2020}]{KormendyHo2013}. The empirical relations are key ingredients in numerical, theoretical, and semi-analytic models to reproduce the galaxy population properties \citep{Croton2006,Schaye2015,Sijacki2015,Volonteri2016,Thomas2019,Li2020}.

Among the relations between BH mass ($M_\bullet$) and host galaxy bulge properties, the BH mass-bulge mass and the BH mass-bulge velocity dispersion ($\sigma_e$)\footnote{We refer to $\sigma_e$ as the stellar velocity dispersion measured inside $R_e$, the bulge effective radius.} correlations stand out due to their equally small scatter ($\sim 0.29\,$dex; \citealt{KormendyHo2013}). Several studies have attempted to find correlations with other galaxy properties (e.g., \citealt{Saglia2016,deNicola2019}), but the uncertainties in the quantities prevent definitive conclusions from being reached. In all cases, the difficulty lies in measuring $M_\bullet$ and $\sigma_e$ for a sufficiently large galaxy population. In the very nearby universe, both quantities can be directly computed. The BH masses can be measured from dynamical analysis for massive BHs (see review by \citealt{KormendyHo2013}), while spatially resolved galaxy spectra are needed to measure $\sigma_e$ (e.g., \citealt{Jorgensen1995,Cappellari2006}). Beyond the nearby universe, instrumental resolution limits dynamical measurements of BH mass. We have to rely on indirect measurements that can only be applied for active galactic nucleus (AGN) host galaxies. Based on the response of the broad-line region (BLR) gas to the variable AGN continuum radiation, the reverberation mapping (RM) technique \citep{Blandford1982,Peterson2004,Kaspi2005,Bentz2013,Shen2015,Bentz2018,Lira2018,Hu2021,Kaspi2021} has proven useful in estimating BH masses. This is possible because nearby RM AGNs follow a $M_\bullet-\sigma_e$ relation that is roughly parallel to that of inactive systems, offering a calibration method \citep{Gebhardt2000b,Ferrarese2001,Nelson2004,Ho2014}. The extension of this technique, the single-epoch virial BH mass estimate \citep{Vestergaard2006,Greene2005,Ho2015}, has allowed statistical studies of BH masses to be carried out for representative samples of the active galaxy population. The advent of high-resolution spectro-astrometric measurements with interferometry \citep[e.g.,][]{Gravity2018,Gravity2023} promises a new option for measuring  $M_\bullet$ in active systems. 

Studying AGNs in the context of $M_\bullet-\sigma_e$ is still a difficult task, mainly because quantifying $\sigma_e$ for such systems is challenging. Stellar velocity dispersion ($\sigma_*$) measurements are susceptible to the underlying stellar population properties, and their relative contribution to the observed luminosity weighted galaxy spectrum. This is probably more significant for AGNs, which have complex stellar populations because of the prevalence of ongoing star formation \citep{Jarvis2020,Shangguan2020b,Torbaniuk2021,Xie2021,Zhuang2022,Molina2023}. In addition, the rich emission-line spectrum of type~1 AGNs blends and confuses with the stellar absorption features of the host galaxy. Moreover, the AGN continuum strongly dilutes the host galaxy spectrum. These problems are specially concerning when measuring $\sigma_*$ from stellar features in the optical \citep{Greene2006b}. Observing in the near-infrared (near-IR) is a plausible alternative (e.g., \citealt{Woo2013}); however, the AGN glare also limits the modeling of the stellar features at those wavelengths, even when adaptive optics-assisted observations are employed \citep{Dasyra2007,Watson2008,Grier2013}. Another option is to model narrow emission lines, such as [O\,{\sc iii}]$\lambda 5007$, to adopt the gas kinematics as a surrogate for $\sigma_*$ (e.g., \citealt{Nelson2000,Shields2003}). Nevertheless, those estimates are prone to large uncertainties \citep{Nelson2000,Onken2004,Bonning2005,Greene2005b,Bennert2018}. To avoid these important complications, the best strategy is to target the Calcium triplet (CaT) absorption lines (e.g., \citealt{Ferrarese2001,Barth2002,Nelson2004,Onken2004,Greene2006,Woo2010}) at 8498, 8542, and 8662\,\r{A}, the ``gold standard'' for deriving stellar velocity dispersion in active galaxies \citep{Greene2006b}. The key advantages of using the CaT for measuring $\sigma_*$ (or $\sigma_e$) is that these stellar features are largely insensitive to the underlying stellar population properties \citep{Dressler1984}, and they are located in a spectral window that is mostly free of strong AGN emission lines. However, observing the CaT for large samples of active galaxies has been prohibitive because the absorption features can be strongly affected by sky emission, and they quickly get redshifted out of the spectral window that is accessible with optical spectrographs, only allowing $\sigma_*$ to be characterized for low-redshift sources.

\begin{figure}
\centering
\includegraphics[width=0.9\columnwidth]{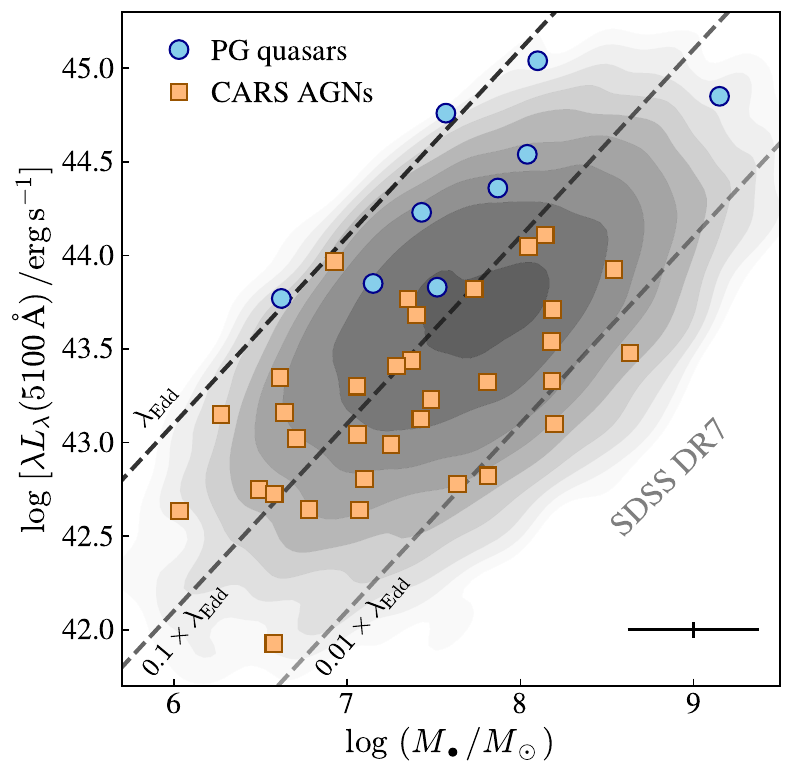}
\caption{\label{fig:sample} AGN monochromatic luminosity at 5100\,\r{A} as a function of the BH mass. The error bar in the bottom right corner represents the typical uncertainty of the quantities. The dashed lines correspond to fixed Eddington ratios. The contours show the distribution of the broad-lined AGNs taken from the 7th data release of the Sloan Digital Sky Survey (SDSS; \citealt{Abolfathi2018}) at $z < 0.35$ presented by \citet{Liu2019}.}
\end{figure}

For this work, we used seeing-limited and ground-layer adaptive optics-aided Very Large Telescope (VLT) Multi Unit Spectroscopic Explorer (MUSE) observations to model the CaT and estimate the bulge velocity dispersion for the host galaxies of local type~1 AGNs and quasars. Our sample was built from the Palomar-Green (PG) quasar survey \citep{Boroson1992} and the Close AGN Reference Survey (CARS; \citealt{Husemann2022}), with all of these at $z \lesssim 0.1$ (Figure~\ref{fig:sample}). We analyzed these systems in the context of the $M_\bullet-\sigma_e$ relation, finding no difference between the active and inactive galaxies. Section~\ref{sec:obs} summarizes the properties of the active galaxy sample and outlines the observations. Section~\ref{sec:main_methods} details the procedures used to derive $M_\bullet$ and $\sigma_e$. Section~\ref{sec:Results} presents the PG quasars and CARS AGNs in the context of the $M_\bullet-\sigma_e$ relation, and examines possible deviations from this relation in terms of the Eddington ratio. We discuss and summarize our findings in Section~\ref{sec:Summ}. Hereinafter, we refer to the central stellar spheroid component of the host galaxy simply as the ``bulge,'' which pertains to both classical and pseudo bulges \citep{Kormendy2004}, unless explicitly stated. We adopted a $\Lambda$CDM cosmology with $\Omega_m = 0.308$, $\Omega_\Lambda = 0.692$, and $H_0 = 67.8\,$km\,s$^{-1}$\,Mpc$^{-1}$ \citep{Planck2016}.

\begin{table*}
        \tiny
        \centering
        \def\arraystretch{1.3}
        \setlength\tabcolsep{3pt}
        \caption{\label{tab:sample} Basic parameters of the sample.}
        \vspace{-0.2mm}
        \begin{tabular}{ccccccccccccc}
                \hline
                \hline
                Object & R.A. & Decl. & $z$ & $D_L$ & Morphology & $\log M_*$ & $\log M_\bullet$ & $\log \lambda L_\lambda(5100\,\mathrm{\AA})$ & \multicolumn{2}{c}{$R_e$} & $n$ & $B/T$\\
                & (J2000.0) & (J2000.0) & & (Mpc) & &($M_\odot$) & ($M_\odot$) & (erg\,s$^{-1}$) & ($\arcsec$) & (kpc) & & \\
                (1) & (2) & (3) & (4) &(5) & (6) & (7) & (8)& (9) & \multicolumn{2}{c}{(10)} & (11) & (12) \\
                \hline
                \multicolumn{13}{c}{Palomar-Green Quasars}\\
                \hline
                PG\,0050+124    & 00:53:34.94 & +12:41:36.2    & 0.0611 & 282 & Disk & 11.12$^{+0.3}_{-0.3}$ & 7.57 & 44.76 & 1.33 & 1.62 & 1.7 & 0.52 \\
                PG\,0923+129    & 09:26:03.29 & +12:44:03.6    & 0.0287 & 131 & Disk & 10.71$^{+0.3}_{-0.3}$ & 7.52 & 43.83 & 1.01 & 0.60 & 1.0 & 0.37 \\
                PG\,0934+013      & 09:37:01.05 & +01:05:43.2    & 0.0506 & 230 & Disk & 10.38$^{+0.3}_{-0.3}$ & 7.15 & 43.85 & 0.57 & 0.58 & 0.5 & 0.05 \\
                PG\,1011$-$040 & 10:14:20.69 & $-$04:18:40.5 & 0.0584 & 268 & Disk & 10.87$^{+0.3}_{-0.3}$ & 7.43 & 44.23 & 0.84 & 0.98 & 2.8 & 0.14 \\
                        PG\,1126$-$041 & 11:29:16.66 & $-$04:24:07.6 & 0.0601 & 278 & Disk & 10.85$^{+0.3}_{-0.3}$ & 7.87 & 44.36 & 0.66 & 0.79 & 4.0 & 0.37 \\
                PG\,1211+143    & 12:11:17.67 & +14:03:13.2    & 0.0815 & 400 & Disk & 10.38$^{+0.3}_{-0.3}$ & 8.10 & 45.04 & 0.95 & 1.51 & 5.8 & 0.72 \\
                PG\,1244+026    & 12:46:35.25 & +02:22:08.8    & 0.0484 & 220 & Disk & 10.19$^{+0.3}_{-0.3}$ & 6.62 & 43.77 & 0.10 & 0.10 & 1.0 & 0.03 \\
                PG\,1426+015    & 14:29:06.57 & +01:17:06.2    & 0.0866 & 401 & Merger & 11.05$^{+0.3}_{-0.3}$ & 9.15 & 44.85 & 0.98 & 1.64 & 4.0 & 1.0\\
                PG\,2130+099    & 21:32:27.81 & +10:08:19.5    & 0.0631 & 292 & Disk & 10.85$^{+0.3}_{-0.3}$ & 8.04 & 44.54 & 2.26 & 2.83 & 4.0 & 0.33 \\
                \hline
                \multicolumn{13}{c}{CARS AGNs}\\
                \hline
                HE\,0021$-$1810 & 00:23:39.34 & $-$17:53:54.6 & 0.0537 & 247 & Spheroidal & 10.64$^{+0.04}_{-0.05}$ & 7.81 & 42.82 & \ldots & \ldots & \ldots & \ldots \\
                HE\,0021$-$1819 & 00:23:55.29 & $-$18:02:51.0 & 0.0533 & 245 & Disk & 10.50$^{+0.04}_{-0.05}$ & 7.10 & 42.80 & 1.9 & 2.04 & 1.4 & 0.33 \\
                HE\,0040$-$1105 & 00:42:36.76 & $-$10:49:22.4 & 0.0419 & 191 & Spheroidal & 10.16$^{+0.13}_{-0.10}$ & 7.43 & 43.13 & 0.9 & 0.77 & 3.5 & 0.5 \\
                HE\,0108$-$4743 & 01:11:09.68 & $-$47:27:35.6 & 0.0239 & 108 & Disk & 9.77$^{+0.18}_{-0.10}$ & 6.03 & 42.63 & 3.0 & 1.49 & 0.8 & 0.94 \\
                HE\,0114$-$0015 & 01:17:03.51 & +00:00:27.9    & 0.0458 & 210 & Disk & 10.47$^{+0.13}_{-0.19}$ & 6.78 & 42.64 & 2.8 & 2.60 & 0.3 & 0.97 \\
                HE\,0119$-$0118 & 01:21:59.76 & $-$01:02:24.4 & 0.0548 & 253 & Disk & 10.91$^{+0.02}_{-0.06}$ & 7.40 & 43.68 & 2.7 & 2.97 & 0.9 & 0.91 \\
                HE\,0203$-$0031$^*$ & 02:06:15.97 & $-$00:17:28.9 & 0.0425 & 194 & Peculiar & 10.88$^{+0.02}_{-0.02}$ & 8.17 & 43.79 & 3.3 & 2.83 & 3.8 & 0.55 \\
                HE\,0212$-$0059 & 02:14:33.58 & $-$00:45:59.9 & 0.0264 & 119 & Disk & 10.59$^{+0.01}_{-0.01}$ & 8.18 & 43.33 & 1.2 & 0.66 & 0.7 & 1.0 \\
                HE\,0224$-$2834 & 02:26:25.70 & $-$28:20:59.2 & 0.0602 & 278 & Peculiar & 10.13$^{+0.19}_{-0.17}$ & 8.18 & 43.54 & 2.6 & 3.12 & 2.6 & 0.17 \\
                HE\,0227$-$0913 & 02:30:05.51 & $-$08:59:53.5 & 0.0165 & 74 & Disk & 9.92$^{+0.17}_{-0.12}$ & 6.27 & 43.15 & 0.7 & 0.24 & 1.0 & 0.96 \\
                HE\,0232$-$0900 & 02:34:37.79 & $-$08:47:15.5 & 0.0427 & 195 & Peculiar & 10.88$^{+0.23}_{-0.12}$ & 8.05 & 44.05 & 3.7 & 3.22 & 1.4 & 0.33 \\
                HE\,0253$-$1641 & 02:56:02.70 & $-$16:29:14.7 & 0.0319 & 145 & Disk & 10.28$^{+0.11}_{-0.24}$ & 6.64 & 43.16 & 3.3 & 2.17 & 0.9 & 0.80 \\
                HE\,0345+0056    & 03:47:40.17 & +01:05:14.8    & 0.0310 & 140 & Spheroidal & 8.85$^{+0.62}_{-0.26}$ & 6.92 & 43.97 & 1.3 & 0.83 & 2.0 & 0.06 \\
                HE\,0351+0240    & 03:54:09.48 & +02:49:31.3    & 0.0354 & 161 & Peculiar & 9.85$^{+0.34}_{-0.70}$ & 7.48 & 43.23 & 1.5 & 1.09 & 5.2 & 0.09 \\
                HE\,0412$-$0803 & 04:14:52.65 & $-$07:55:40.4 & 0.0380 & 173 & Spheroidal & 10.08$^{+0.10}_{-0.12}$ & 7.73 & 43.82 & 1.7 & 1.32 & 1.8 & 0.33 \\
                HE\,0429$-$0247 & 04:31:37.12 & $-$02:41:23.4 & 0.0423 & 193 & Uncertain & 9.18$^{+0.51}_{-0.10}$ & 6.61 & 43.35 & 0.9 & 0.78 & 0.7 & 0.86 \\
                HE\,0433$-$1028 & 04:36:22.30 & $-$10:22:33.1 & 0.0355 & 161 & Disk & 10.80$^{+0.08}_{-0.09}$ & 7.35 & 43.77 & 10.1 & 7.37 & 0.2 & 0.83 \\
                HE\,0853+0102    & 08:55:54.25 & +00:51:11.9     & 0.0526 & 241 & Disk & 10.54$^{+0.04}_{-0.10}$ & 7.25 & 42.99 & 0.9 & 1.07 & 0.8 & 0.86 \\
                HE\,0949$-$0122 & 09:52:19.05 & $-$01:36:43.7 & 0.0197 & 88 & Spheroidal & 10.02$^{+0.09}_{-0.31}$ & 6.58 & 42.72 & 1.5 & 0.62 & 3.5 & 0.05 \\
                HE\,1017$-$0305 & 10:19:32.84 & $-$03:20:14.6 & 0.0491 & 225 & Disk & 10.93$^{+0.10}_{-0.15}$ & 8.18 & 43.71 & 3.8 & 3.77 & 2.2 & 0.28 \\
                HE\,1029$-$1831 & 10:31:57.33 & $-$18:46:32.7 & 0.0405 & 185 & Disk & 10.49$^{+0.06}_{-0.18}$ & 7.06 & 43.30 & 2.1 & 1.74 & 0.7 & 0.50 \\
                HE\,1107$-$0813 & 11:09:48.50 & $-$08:30:14.7 & 0.0585 & 270 & Disk & 11.17$^{+0.20}_{-0.38}$ & 8.14 & 44.11 & 1.4 & 1.64 & 3.0 & 0.39 \\
                HE\,1108$-$2813 & 11:10:47.99 & $-$28:30:04.1 & 0.0240 & 108 & Disk & 10.29$^{+0.11}_{-0.05}$ & 7.06 & 43.04 & 5.8 & 2.90 & 0.2 & 0.99 \\
                HE\,1237$-$0504 & 12:39:39.42 & $-$05:20:38.5 & 0.0083 & 37 & Disk & 10.92$^{+0.01}_{-0.01}$ & 7.07 & 42.64 & 5.5 & 0.97 & 1.5 & 0.97 \\
                HE\,1248$-$1356 & 12:51:32.39 & $-$14:13:16.1 & 0.0145 & 65 & Disk & 10.31$^{+0.01}_{-0.01}$ & 6.58 & 41.93 & 1.1 & 0.34 & 1.5 & 0.99 \\
                HE\,1330$-$1013 & 13:32:39.15 & $-$10:28:52.4 & 0.0225 & 101 & Disk & 10.69$^{+0.03}_{-0.13}$ & 6.49 & 42.75 & 8.3 & 3.90 & 1.8 & 0.44 \\
                HE\,1353$-$1917 & 13:56:36.77 & $-$19:31:44.9 & 0.0348 & 158 & Disk & 10.99$^{+0.03}_{-0.06}$ & 8.20 & 43.10 & 11.4 & 8.16 & 2.7 & 0.01 \\
                HE\,1417$-$0909 & 14:20:06.34 & $-$09:23:12.8 & 0.0437 & 200 & Disk & 10.23$^{+0.10}_{-0.17}$ & 7.37 & 43.44 & 2.2 & 1.96 & 0.6 & 1.0 \\
                HE\,2128$-$0221 & 21:30:49.96 & $-$02:08:14.0 & 0.0527 & 242 & Disk & 9.93$^{+0.21}_{-0.67}$ & 6.71 & 43.02 & 2.1 & 2.23 & 1.7 & 0.01 \\
                HE\,2211$-$3903 & 22:14:42.06 & $-$38:48:22.7 & 0.0397 & 181 & Disk & 9.83$^{+0.22}_{-0.21}$ & 7.81 & 43.32 & 11.0 & 8.93 & 0.2 & 0.94 \\
                HE\,2222$-$0026 & 22:24:35.22 & $-$00:11:03.4 & 0.0581 & 268 & Disk & 10.20$^{+0.09}_{-0.15}$ & 7.28 & 43.41 & 1.1 & 1.28 & 1.1 & 0.50 \\
                HE\,2233+0124    & 22:35:41.94 & +01:39:34.5    & 0.0567 & 262 & Disk & 10.71$^{+0.10}_{-0.03}$ & 8.63 & 43.48 & 2.5 & 2.84 & 8.3 & 0.44 \\
                HE\,2302$-$0857 & 23:04:43.70 & $-$08:41:08.3 & 0.0470 & 215 & Disk & 11.20$^{+0.09}_{-0.13}$ & 8.54 & 43.93 & 3.9 & 3.72 & 5.5 & 0.39 \\
                \hline
                \end{tabular}
                \tablefoot{ (1) Source name. (2) Right ascension. (3) Declination. (4) Redshift. (5) Luminosity distance. (6) Morphological type of the host galaxy. For the PG quasars, we adopt the results of \citet{Kim2017}, \citet{Zhang2016}, and \citet{Zhao2021}; for the CARS AGNs, we follow \citet{Husemann2022}, but we relabel their ``irregular" category to ``peculiar" to avoid confusion with the ``dwarf irregular" systems. We also relabel their ``bulge-dominated'' category to simply ``spheroidal.'' (7) Host galaxy stellar mass \citep{Zhao2021,Smirnova-Pinchukova2022}. (8) Black hole mass, based on the calibration of \citet{Ho2015}; the $1\sigma$ uncertainty is $\sim 0.32-0.38\,$dex (Section~\ref{sec:BH_mass}). (9) AGN monochromatic luminosity at 5100\,\r{A}; for the CARS data, the values were derived from the broad H$\beta$ luminosity following \citet{Greene2005}. (10) Bulge effective radius. (11) Bulge \cite{Sersic1963} index. (12) Bulge-to-total ($B/T$) light fraction. The bulge properties for the PG quasars are obtained from optical and near-IR HST image modeling \citep{Veilleux2009,Kim2017,Zhao2021}, while for the CARS AGNs we adopt the MUSE-based $i-$band image models of \citet{Husemann2022}, assuming the S\'ersic model subcomponent with smaller $R_e$ value. (*) For this target, the AGN optical spectrum and bulge properties were derived by \citet{Bennert2011} and \citet{Bennert2021}, respectively.}
\end{table*}

\begin{table}
        \small
        \centering      
        \def\arraystretch{1.3}
        \setlength\tabcolsep{2pt}
        \caption{\label{tab:obssetup} MUSE observational setup.}
        \vspace{0.2mm}
        \begin{tabular}{ccccc}
                \hline
                \hline
                Object & Observation & Exposure  & Image & Instrument\\
                & Date & Time\,(s) & Quality\,($\arcsec$) & Mode\\
                (1) & (2) & (3) & (4) & (5)\\
                \hline
                \multicolumn{5}{c}{Palomar-Green Quasars}\\
                \hline
                PG\,0050+124 & 06\,Oct.\,2019 & 9180 & 0.83 & WFM-AO\\
                PG\,0923+129 & 28\,Apr.\,2019 & 2440 & 1.49 & WFM-AO\\
                PG\,0934+013 & 04\,Apr.\,2015 & 1350 & 0.81 & WFM-noAO\\
                PG\,1011$-$040 & 15\,Jun.\,2015 & 600 & 1.08 & WFM-noAO\\
                PG\,1126$-$041 & 10\,Sep.\,2015 & 900 & 1.01 & WFM-noAO\\
                        PG\,1211+143 & 01\,Apr.\,2016 & 2800 & 0.76 & WFM-noAO\\
                        PG\,1244+026 & 30\,May\,2019 & 2440 & 1.08 & WFM-AO\\
                PG\,1426+015 & 04\,Apr.\,2016 & 2800 & 0.60 & WFM-noAO\\
                PG\,2130+099 & 25\,Sep.\,2019 & 2440 & 0.70 & WFM-AO\\
                \hline
                \multicolumn{5}{c}{CARS AGNs}\\
                \hline
                HE\,0021$-$1810 & 19\,Apr.\,2016 & 1400 & 0.94 & WFM-noAO \\ 
                HE\,0021$-$1819 & 19\,Apr.\,2016 & 1400 & 1.05 & WFM-noAO \\ 
                HE\,0040$-$1105 & 19\,Apr.\,2016 & 800 & 1.41 & WFM-noAO \\ 
                HE\,0108$-$4743 & 19\,Apr.\,2016 & 600 & 1.27 & WFM-noAO \\ 
                HE\,0114$-$0015 & 12\,Jul.\,2016 & 900 & 0.66 & WFM-noAO \\ 
                HE\,0119$-$0118 & 19\,Apr.\,2016 & 600 & 0.95 & WFM-noAO \\ 
                HE\,0203$-$0031 & 27\,Apr.\,2016 & 800 & 1.12 & WFM-noAO \\ 
                HE\,0212$-$0059 & 12\,Dec.\,2017 & 2200 & 0.95 & WFM-noAO \\ 
                HE\,0224$-$2834 & 06\,Jul.\,2016 & 900 & 1.35 & WFM-noAO \\ 
                HE\,0227$-$0913 & 27\,Apr.\,2016 & 1200 & 1.41 & WFM-noAO \\ 
                HE\,0232$-$0900 & 19\,Apr.\,2016 & 600 & 1.03 & WFM-noAO \\ 
                HE\,0253$-$1641 & 19\,Apr.\,2016 & 800 & 0.83 & WFM-noAO \\ 
                HE\,0345+0056 & 19\,Apr.\,2016 & 1600 & 0.86 & WFM-noAO \\ 
                HE\,0351+0240 & 19\,Apr.\,2016 & 1600 & 0.72 & WFM-noAO \\ 
                HE\,0412$-$0803 & 19\,Apr.\,2016 & 1600 & 0.88 & WFM-noAO \\ 
                HE\,0429$-$0247 & 19\,Apr.\,2016 & 2400 & 0.81 & WFM-noAO \\ 
                HE\,0433$-$1028 & 19\,Apr.\,2016 & 600 & 0.55 & WFM-noAO \\ 
                HE\,0853+0102 & 19\,Apr.\,2016 & 1200 & 0.57 & WFM-noAO \\
                HE\,0949$-$0122 & 11\,Mar.\,2021 & 2300 & 1.16 & WFM-noAO \\  
                HE\,1017$-$0305 & 23\,Jun.\,2016 & 900 & 1.12 & WFM-noAO \\ 
                HE\,1029$-$1831 & 19\,Apr.\,2016 & 600 & 0.66 & WFM-noAO \\ 
                HE\,1107$-$0813 & 27\,Jun.\,2016 & 1350 & 0.90 & WFM-noAO \\ 
                HE\,1108$-$2813 & 19\,Apr.\,2016 & 400 & 0.49 & WFM-noAO \\ 
                HE\,1237$-$0504 & 24\,May\,2017 & 3600 & 0.52 & WFM-noAO \\ 
                HE\,1248$-$1356 & 28\,Jun.\,2016 & 600 & 0.57 & WFM-noAO \\ 
                HE\,1330$-$1013 & 20\,Jun.\,2016 & 600 & 0.67 & WFM-noAO \\ 
                HE\,1353$-$1917 & 20\,Jun.\,2016 & 900 & 0.71 & WFM-noAO \\ 
                HE\,1417$-$0909 & 20\,Jun.\,2016 & 1350 & 0.68 & WFM-noAO \\ 
                HE\,2128$-$0221 & 25\,May\,2016 & 1350 & 0.68 & WFM-noAO \\ 
                HE\,2211$-$3903 & 06\,Jun.\,2016 & 900 & 0.57 & WFM-noAO \\ 
                HE\,2222$-$0026 & 20\,Jun.\,2016 & 1350 & 0.64 & WFM-noAO \\ 
                HE\,2233+0124 & 23\,Jun.\,2016 & 1350 & 1.00 & WFM-noAO \\ 
                HE\,2302$-$0857 & 09\,Jun.\,2016 & 600 & 0.66 & WFM-noAO \\ 
        \hline
        \end{tabular}
\tablefoot{(1) Source name. (2) Date of the MUSE observations. (3) Total exposure time. (4) Image quality of the observations as reported in the data cube header keyword ``SKY\_RES.'' (5) MUSE instrument mode. Most of the observations correspond to wide-field-mode natural-seeing (WFM-noAO), but four were aided by ground-layer adaptive optics (WFM-AO).}
\end{table}

\section{Sample and observations}
\label{sec:obs}

We use VLT-MUSE wide-field-mode integral field unit (IFU) observations for our sample, which consists of 42 local ($z < 0.1$; Table~\ref{tab:sample}) type~1 AGNs, nine PG quasars \citep{Boroson1992} and 33 CARS AGNs \citep{Husemann2022}. All observations were conducted as part of multiple European Southern Observatory (ESO) programs [094.B$-$0345(A), 095.B$-$0015(A), 097.B$-$0080(A), 099.B$-$0242(B), 099.B$-$0294(A), 0101.B$-$0368(B), 0103.B$-$0496, 0104.B$-$0151(A), and 106.21C7.002], spanning June 2015 to March 2021. In a single snapshot, MUSE provides a wide field-of-view of approximately $1\arcmin \times 1\arcmin$, with a $0\farcs2 \times 0\farcs2$ pixel size. It generates $\sim 90,000$ spectra per pointing. The wavelength coverage of MUSE ranges from $\sim$4700 to $\sim$9350\,\r{A}, with a wavelength sampling of 1.25\,\r{A}\,channel$^{-1}$ at a mean resolution of $R \approx 3000$ (full width at half maximum FWHM\,$= 2.65$\,\r{A}). We analyze archival MUSE data cubes obtained from the ESO science portal.\footnote{http://archive.eso.org/scienceportal/home} These data cubes correspond to a combination of multiple observing blocks, typically ranging from two to four per target. In the particular case of PG\,0050+124, we used the ``MUSE-DEEP'' data cube,\footnote{http://www.eso.org/rm/api/v1/public/releaseDescriptions/102} which is a combination of 15 single observing blocks  aimed at maximizing the signal contrast \citep{Weilbacher2020}. Most of the MUSE observations were conducted under natural-seeing conditions, with typical values $\sim 0\farcs5-1\farcs4$ (Table~\ref{tab:obssetup}). Only four observations were assisted by the ground-layer adaptive optics module. In these cases, the MUSE spectra were masked at the 5840--5940\,\r{A} wavelength range surrounding the Na~I~D$\lambda \lambda 5890,5896$ emission by the data reduction pipeline.

We further applied the Zurich Atmosphere Purge (ZAP) sky-subtraction tool (version ``2.1.dev''; \citealt{Soto2016}) to remove residual sky features. We set \textsc{cfwidthSP}\,$= 5$, while keeping the other parameters at their default values. Additionally, we masked all spectra at wavelengths where strong sky-line residuals are present in the preprocessed data cubes (5578.5, 5894.6, 6301.7, 6362.5, and 7640\,\r{A}). To account for instrumental resolution, we adopt the MUSE line-spread function parameterization of \citet{Guerou2017} to correct the line widths.

\subsection{PG quasars}
\label{sec:PGQuasars}
We use the available MUSE data for nine local quasar host galaxies previously presented in \citet{Molina2022}. These systems are member of the broader sample of 87 $z < 0.5$ quasars \citep{Boroson1992} belonging to the PG survey of optical/ultraviolet color-selected quasars \citep{Schmidt1983}. These targets have substantial multiwavelength data across the entire electromagnetic spectrum, ranging from X-ray \citep{Reeves2000,Bianchi2009} to optical \citep{Boroson1992,Ho2009}, mid-IR \citep{Shi2014,Xie2021,Xie2022}, far-IR \citep{Petric2015,Shangguan2018,Zhuang2018}, mm \citep{Shangguan2020,Shangguan2020b,Molina2021}, and radio \citep{Kellermann1989,Kellermann1994,Silpa2020,Silpa2023} wavelengths. High-resolution ($\sim 0\farcs1$) \textit{Hubble Space Telescope} (HST) imaging in the optical and near-IR are available for many of the host galaxies, securing accurate characterization of the host galaxy morphology \citep{Kim2008,Zhang2016,Kim2019,Zhao2021}. Stellar masses ($M_*$) were computed from the multiband host galaxy images after applying the mass-to-light ratios of \citet[see \citealt{Zhao2021} for more details]{Bell2001}. We estimate bolometric luminosities ($L_{\rm bol}$) from the AGN monochromatic luminosity at 5100\,\r{A}, $\lambda L_\lambda(5100\,\mathrm{\AA})$, by adopting the conversion of \citet{Richards2006}, $L_{\rm bol} = 10\, \lambda L_\lambda (5100\, \mathrm{\AA})$. Our PG quasar subsample is characterized by $\langle L_{\rm bol} \rangle = 10^{45.6}\,$erg\,s$^{-1}$, $\langle M_\bullet \rangle = 10^{8.3}\,M_\odot$, $\langle M_* \rangle = 10^{10.8}\,M_\odot$, and $\langle z \rangle = 0.060$.

\subsection{CARS AGNs}
\label{sec:CARS}
The CARS sample was selected from the broader Hamburg/ESO survey (HES) of ultraviolet-excess sources covering an area of $\sim 9000\, \rm deg^2$ in the southern hemisphere. The type~1 AGNs were confirmed through follow-up spectroscopy \citep{Wisotzki2000,Schulze2009}. Specifically, CARS corresponds to a representative survey of 41 systems randomly selected from the broader subsample of 99 HES objects at $z \lesssim 0.06$ \citep{Bertram2007}. CARS probes the bright tail of the AGN luminosity function in the local universe \citep{Schulze2009}. The CARS host galaxies have been observed at multiple wavelengths, including the mm \citep{Bertram2007} and radio \citep{Konig2009}. The panchromatic spectral energy distribution decomposition and stellar mass estimation for the host galaxies are presented in detail in \citet{Smirnova-Pinchukova2022}. Available IFU data are presented in \citet{Husemann2022} for 41 targets, 37 of these observed by MUSE. From the MUSE dataset, we discard two targets that do not correspond to type~1 AGNs, but instead starbursts with extremely blue continuum and broad-line components tracing starburst-driven outflows \citep{Husemann2022}. Another two targets overlap with the PG quasar sample and were also discarded. Hence, we analyze 33 CARS systems (Table~\ref{tab:sample}). These CARS host galaxies are characterized by $\langle z \rangle = 0.042$, $\langle L_{\rm bol} \rangle = 10^{44.5}\,$erg\,s$^{-1}$, $\langle M_\bullet \rangle = 10^{7.9}\,M_\odot$, and $\langle M_* \rangle = 10^{10.6}\,M_\odot$.\\
\\

\section{Methods}
\label{sec:main_methods}

Our aim is to study local type~1 AGNs and quasars in the context of the $M_\bullet$--$\sigma_e$ relation. To characterize $\sigma_e$, we focus on modeling the CaT. We minimize the effect of the AGN emission in diluting the stellar spectrum features by using annular (elliptical) apertures when extracting the spectra. For each source, we measure $\sigma_*$ as close as possible to the bulge half-light radius ($R_e$). We develop an annular aperture correction recipe to estimate $\sigma_e$ from $\sigma_*$. We further correct for systematics associated with MUSE instrumental artifacts to ensure accurate estimates. Additionally, we control for AGN emission diluting the stellar continuum and the signal-to-noise (S/N) of the stellar features. 

\subsection{Host galaxy spectra extraction} 
\label{sec:sp_extract}

We begin by characterizing the projected geometry of the host galaxies on the sky. We collapse the MUSE data cubes across the spectral axis to derive ``white-light'' images. For each white-light image, we use the \textsc{background2D} task from \textsc{photutils} \citep{Bradley2022} to compute the background level. Then, we smooth the white-light image with a $1\arcsec$-wide Gaussian kernel and apply \textsc{detect\_sources} to build a segmentation map. During this process, we mask sources with S/N\,$<5$. We use the segmentation map to isolate the host galaxy from any other sources in the unsmoothed white-light image, and we apply \textsc{SourceCatalog} to derive the system position angle and ellipticity. The AGN location is assumed to coincide with the center of the host galaxy. We use the geometric parameters to construct a series of concentric annuli with thickness equal to the point-spread function (PSF) FWHM. The spectra are bounded to a window of $\pm1000\,$km\,s$^{-1}$ to enclose the redshifted CaT,\footnote{For PG\,1426+015 we use a window of width $\pm 1500$\,km\,s$^{-1}$ to better characterize the continuum.} with a wavelength upper limit set to 9300\,\r{A} in the observer frame to avoid the red spectral limit of MUSE. The spectra are masked whenever sky lines are present. These sky lines are identified in the corresponding variance spectra, analogously extracted from the MUSE variance data cubes. Nonetheless, we recall that the sky-line features have been minimized by applying the ZAP tool on the data cubes.

\subsection{MUSE instrumental feature correction} 
\label{sec:sp_correction}

\begin{figure}[!h]
\centering
\includegraphics[width=0.9\columnwidth]{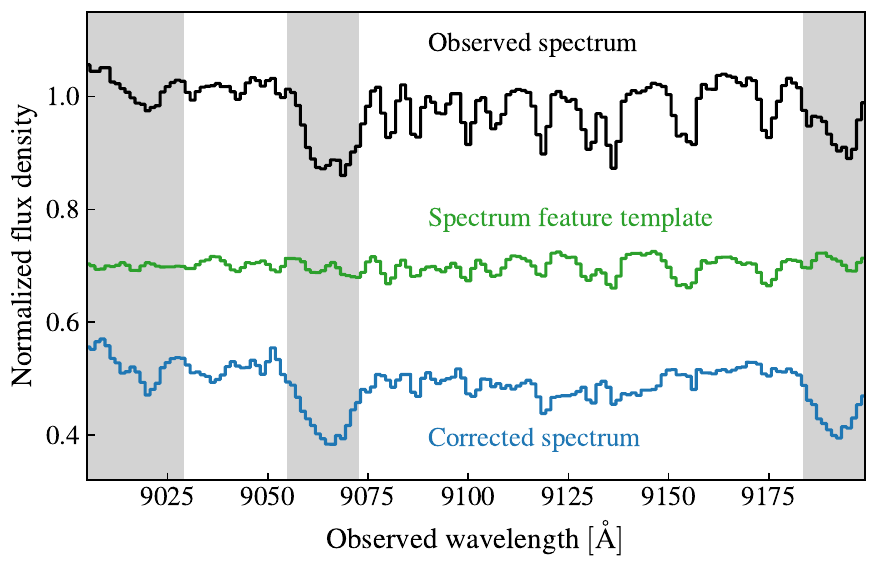}
\caption{\label{fig:sp_feature} Example of MUSE instrumental feature correction. The shaded regions highlight the masked wavelengths that encompass the CaT. The spectra have been vertically shifted to improve figure visualization.}
\end{figure}

We identify a MUSE instrumental feature at 9060--9180\,\r{A} in the observer frame (Figure~\ref{fig:sp_feature}), which must be corrected. The key characteristic of this instrumental feature is its consistent presence at the same wavelength range whenever MUSE observes a relatively bright source, such as a bright field star or the AGN. We speculate that this instrumental feature is caused by minor variations in the instrument sensitivity at specific wavelengths. We employ a ``flat field-like'' procedure to correct for this feature. We note that this instrumental feature overlaps with at least one CaT absorption line for AGNs at $0.046 < z < 0.080$. 

\begin{figure}[!h]
\centering
\includegraphics[width=0.9\columnwidth]{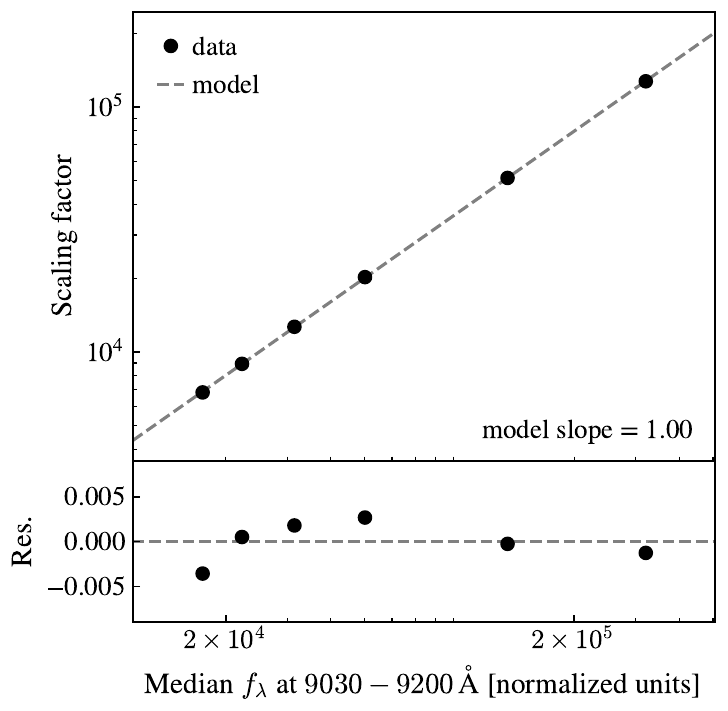}
\caption{\label{fig:normalization_constant} Scaling factor for the instrumental feature template as a function of the median flux density for the annuli-extracted spectra over the 9030--9200\,\r{A} wavelength range (observer frame). The model corresponds to a linear function adjusted to the data. The bottom panel shows the model residuals. We only show the data for one target. We present the rest of the sample in Appendix~\ref{app:IF_charac}.}
\end{figure}

\begin{table}[!h]
        \tiny
        \centering      
        \def\arraystretch{1.2}
        \setlength\tabcolsep{2pt}
        \caption{\label{tab:derived_quantities} Measured and derived quantities.}
        \vspace{0.2mm}
        \begin{tabular}{ccccccccc}
                \hline
                \hline
                Object & AGN & \multicolumn{2}{c}{$R_{\rm ap}$} & EW & S/N & $\sigma_*$  & $F_{\rm ap}$ & $\sigma_e$ \\
                & Subtraction & ($\arcsec$) & (kpc) & (\r{A})& & (km\,s$^{-1}$) & & (km\,s$^{-1}$) \\
                (1) & (2) & \multicolumn{2}{c}{(3)} & (4) & (5) & (6) & (7) & (8) \\
                \hline
                \multicolumn{9}{c}{Palomar-Green Quasars}\\
                \hline          
                PG$\,$0050+124 & Yes & 1.25 & 1.52 & 0.4 & 9.1 & $177\pm9$ & 0.95 & 186$\pm$22 \\
                PG$\,$0923+129 & No & 2.24 & 1.33 & 2.6 & 19.8 & $152\pm2$ & 0.82 & 185$\pm$22 \\
                PG$\,$0934+013 & Yes & 1.44 & 1.47 & 1.2 & 8.5 & $134\pm10$ & 0.73 & 184$\pm$23 \\
                PG$\,$1011$-$040 & Yes & 1.74 & 2.03 & 2.3 & 9.9 & $136\pm6$ & 0.79 & 173$\pm$23 \\
                PG$\,$1126$-$041 & Yes & 2.07 & 2.48 & 1.4 & 10.0 & $175\pm24$ & 0.64 & 274$\pm$43 \\
                PG$\,$1211+143 & Yes & 1.65 & 2.62 & 0.8 & 11.4 & $65\pm10$ & 0.66 & 98$\pm$19 \\
                PG$\,$1244+026$^*$ & Yes & 1.61 & 1.58 & 1.2 & 12.7 & $80\pm7$ & \ldots & 99$\pm$9 \\
                PG$\,$1426+015 & Yes & 1.09 & 1.83 & 0.8 & 4.0 & $208\pm37$ & 0.83 & 250$\pm$53 \\
                PG$\,$2130+099 & Yes & 1.76 & 2.21 & 1.4 & 6.8 & $169\pm8$ & 0.88 & 192$\pm$23 \\          
                \hline
                \multicolumn{9}{c}{CARS AGNs}\\
                \hline
                HE$\,$0021$-$1810$^*$ &No & 0.60 & 0.65 & 1.8 & 6.1 & $139\pm9$ & \ldots & 140$\pm$9 \\
                HE$\,$0021$-$1819 &No & 0.71 & 0.76 & 1.9 & 5.4 & $110\pm9$ & 1.07 & 103$\pm$10 \\
                HE$\,$0040$-$1105 &No & 0.85 & 0.73 & 1.7 & 5.9 & $82\pm9$ & 0.96 & 85$\pm$12 \\
                HE$\,$0108$-$4743 &No & 2.45 & 1.22 & 2.0 & 15.8 & $65\pm2$ & 0.99 & 65$\pm$6 \\
                HE$\,$0114$-$0015 &No & 1.24 & 1.16 & 2.8 & 7.0 & $129\pm7$ & 1.06 & 122$\pm$10 \\
                HE$\,$0119$-$0118 &No & 1.86 & 2.05 & 1.8 & 8.9 & $62\pm5$ & 1.00 & 62$\pm$7 \\
                HE$\,$0203$-$0031 &No & 0.76 & 0.65 & 2.2 & 9.8 & $222\pm7$ & 1.06 & 209$\pm$13 \\
                HE$\,$0212$-$0059 &No & 1.74 & 0.95 & 2.3 & 19.7 & $164\pm3$ & 0.88 & 187$\pm$23 \\
                HE$\,$0224$-$2834 &Yes & 0.89 & 1.06 & 1.3 & 1.3 & 194$\pm$101 & 1.07 & 182$\pm$96 \\
                HE$\,$0227$-$0913$^*$ &Yes & 2.61 & 0.91 & 1.3 & 20.3 & $95\pm3$ & \ldots & 96$\pm$12 \\
                HE$\,$0232$-$0900 &Yes & \ldots & \ldots & \ldots & \ldots & \ldots & \ldots & \ldots \\
                HE$\,$0253$-$1641 &Yes & 1.56 & 1.03 & 1.1 & 13.3 & $96\pm5$ & 1.05 & 91$\pm$8 \\
                HE$\,$0345+0056 &Yes & 1.58 & 1.01 & 0.3 & 14.6 & $80\pm5$ & 0.91 & 87$\pm$11 \\
                HE$\,$0351+0240 &No & 1.31 & 0.95 & 1.5 & 4.4 & $156\pm18$ & 0.85 & 183$\pm$25 \\
                HE$\,$0412$-$0803 &Yes & \ldots & \ldots & \ldots & \ldots & \ldots & \ldots & \ldots \\
                HE$\,$0429$-$0247 &No & 1.83 & 1.58 & 1.4 & 10.2 & $83\pm5$ & 0.79 & 105$\pm$15 \\
                HE$\,$0433$-$1028 &No & 1.11 & 0.81 & 0.6 & 3.2 & $99\pm17$ & 1.18 & 84$\pm$16 \\
                HE$\,$0853+0102 &Yes & 1.14 & 1.21 & 2.6 & 5.7 & $141\pm9$ & 0.92 & 154$\pm$20 \\
                HE$\,$0949$-$0122 &No & 2.49 & 1.03 & 1.0 & 13.9 & $97\pm3$ & 0.77 & 125$\pm$16 \\
                HE$\,$1017$-$0305 &No & 0.71 & 0.70 & 1.3 & 3.9 & $163\pm43$ & 1.10 & 148$\pm$39 \\
                HE$\,$1029$-$1831 &No & 1.32 & 1.09 & 2.0 & 7.6 & $120\pm7$ & 1.02 & 118$\pm$11 \\
                HE$\,$1107$-$0813 &Yes & 1.78 & 2.09 & 1.3 & 10.3 & $169\pm7$ & 0.86 & 196$\pm$23 \\
                HE$\,$1108$-$2813 &No & 1.68 & 0.84 & 1.9 & 10.5 & $108\pm4$ & 1.10 & 98$\pm$9 \\
                HE$\,$1237$-$0504  &No & 5.55 & 0.98 & 2.6 & 20.3 & $155\pm2$ & 0.91 & 170$\pm$19 \\
                HE$\,$1248$-$1356 &No & 1.12 & 0.34 & 1.7 & 12.2 & $88\pm3$ & 0.94 & 94$\pm$10 \\
                HE$\,$1330$-$1013 &No & 2.55 & 1.20 & 2.5 & 7.7 & $116\pm5$ & 1.07 & 108$\pm$10 \\
                HE$\,$1353$-$1917 &No & 1.45 & 1.04 & 1.1 & 3.3 & 160$\pm$68 & 1.10 & 146$\pm$63 \\
                HE$\,$1417$-$0909 &No & 1.59 & 1.41 & 2.0 & 9.3 & 81$\pm$6 & 0.99 & 81$\pm$9 \\
                HE$\,$2128$-$0221 &No & 1.41 & 1.50 & 2.0 & 5.3 & $60\pm13$ & 0.99 & 60$\pm$14 \\
                HE$\,$2211$-$3903 &No & 1.24 & 1.01 & 1.8 & 11.0 & $126\pm6$ & 1.18 & 106$\pm$9 \\
                HE$\,$2222$-$0026 &No & 0.47 & 0.55 & 1.6 & 5.7 & $114\pm10$ & 1.06 & 108$\pm$11 \\
                HE$\,$2233+0124 &No & 0.68 & 0.77 & 2.2 & 5.8 & $166\pm11$ & 0.96 & 173$\pm$14 \\
                HE$\,$2302$-$0857 &Yes & 1.29 & 1.23 & 0.8 & 5.7 & $228\pm22$ & 0.97 & 234$\pm$27 \\
        \hline
        \end{tabular}
\tablefoot{(1) Source name. (2) Whether AGN emission was subtracted before obtaining $\sigma_*$ (Section~\ref{sec:CaT_fitting}). (3) Aperture radius at which the stellar velocity dispersion is extracted. (4) Equivalent width of the Ca\,{\sc ii}\,$\lambda8542$ feature. (5) S/N of the Ca\,{\sc ii}\,$\lambda8542$ feature. (6) Stellar velocity dispersion. (7) Annular aperture correction factor. (8) Stellar velocity dispersion of the bulge. ($^*$) The value of $\sigma_e$ is estimated by applying the average $F_{\rm ap}$ value of the corresponding AGN subsample.}
\end{table}

We characterize the shape of the MUSE instrumental feature by using field stars (identified by \textit{Gaia}) within the MUSE field-of-view. Out of all data cubes, we find that the spectra of 29 bright stars exhibit instrumental features. Stars are seen as point-like sources in the MUSE white-light images, so we model them using a \cite{Moffat1969} function. We extract the stellar spectrum by spatially collapsing the data cube over a circular aperture equal to twice the Moffat model FWHM. Each stellar spectrum is fitted within the 4750--9280\,\r{A} wavelength range using the penalized pixel-fitting ({\tt pPXF}; \citealt{Cappellari2004}) routine. We identify at least ten field stars with good-quality spectrum models, but we only use the four\footnote{These stars correspond to Gaia\,DR3\,3195905922532342656, 2684962933533311488, 2504238273149858816, and 2366885700060472576, which are consistent with being K-type, but were not previously selected by stellar classification.} that deliver the best removal of the MUSE instrumental feature. We use the stellar spectra to obtain an instrumental feature template from the data minus model residuals. Then, we flux-density normalize these templates before proceeding with the MUSE instrumental feature correction. We found that the normalization value for the instrumental feature template slightly varies for each MUSE observation (see Appendix~\ref{app:IF_charac} for more details). Therefore, we determine a specific normalization value for each case. For each target, we fit the spectra extracted from all annuli over the 9030--9200\,\r{A} wavelength range using a power-law continuum model plus the instrumental feature template multiplied by a scaling factor as free parameter. During the fit, we mask a spectral window of $\pm 300$\,km\,s$^{-1}$ around each CaT absorption line, if present.\footnote{For PG\,1011$-$040 we use a window of width $\pm 600$\,km\,s$^{-1}$ because of the presence of calcium in emission \citep{Persson1988}.}  We compare the best-fit scaling factor with the local median spectrum flux density (computed over 9030--9200\,\r{A}), finding that the ratio between these two quantities remains approximately constant for each annulus-extracted spectrum (Figure~\ref{fig:normalization_constant}). Thus, for each target, we calculate the normalization constant by averaging the ratio over all annuli-extracted spectra. Finally, we divide each annulus-extracted spectrum by its corresponding normalized instrumental feature template. Considering the ten field stars with good-quality spectrum models for building different instrumental feature templates, we estimate systematic uncertainty of 0.08\,dex induced by template (star) selection.

\subsection{CaT modeling and stellar velocity dispersion measurements} 
\label{sec:CaT_fitting}

We employ {\tt pPXF} with input stellar templates taken from the INDO-U.S. stellar spectral library of \cite{Valdes2004} to model the spectra. The stellar templates cover the 3460--9464\,\r{A} wavelength range at a uniform spectral resolution of ${\rm FWHM} = 1.35\,$\r{A} \citep{Beifiori2011}. We use a combination of late-type (F, G, K, and M) red giant (luminosity class III) stars of near-solar metallicity, plus an A-type dwarf (luminosity class V) star. We broaden the stellar templates in velocity space, taking into account their spectral resolution difference (in quadrature) with respect to the width of the MUSE line-spread function. We note that our results are not particularly sensitive to our choice of stellar template library, with an uncertainty of 0.04\,dex associated with adopting other options for stellar spectral library (e.g., \citealt{Vazdekis2012}), or a simpler weighted linear combination of A and K stellar templates \citep{Kong2018}. We avoid employing high-order moments (e.g., $h_3$ and $h_4$) when modeling the CaT because of the modest absorption line S/N and data spectral resolution. Even though the contrast between the AGN and its host galaxy is minimized at the CaT wavelengths, the featureless spectrum of the AGN continuum can still dominate the observed emission. The wing of the O\,{\sc i}\,$\lambda 8446$ broad emission line can also interfere in the bluer end of the spectral range considered here \citep{Caglar2020}. To model such possible spectral subcomponents, we include an additive polynomial of order 0 and a multiplicative polynomial of order 3 in the {\tt pPXF} setup. The inclusion of additive and multiplicative polynomials during the fit also helps to correct imperfect sky subtraction or scattered light with the former, and inaccuracies in spectral calibration or mismatches in dust reddening correction with the latter \citep{Cappellari2017}. We caution that the inclusion of an additive polynomial tends to downweight the young (age $\lesssim 1\,$Gyr) stellar templates during the fit; however, the Ca\,{\sc ii} absorption line widths are largely insensitive to stellar population properties \citep{Dressler1984}. Besides $\sigma_*$, deriving other parameters for the host galaxy stellar component is beyond the scope of this work. 

\begin{figure}[!t]
\centering
\includegraphics[width=0.9\columnwidth]{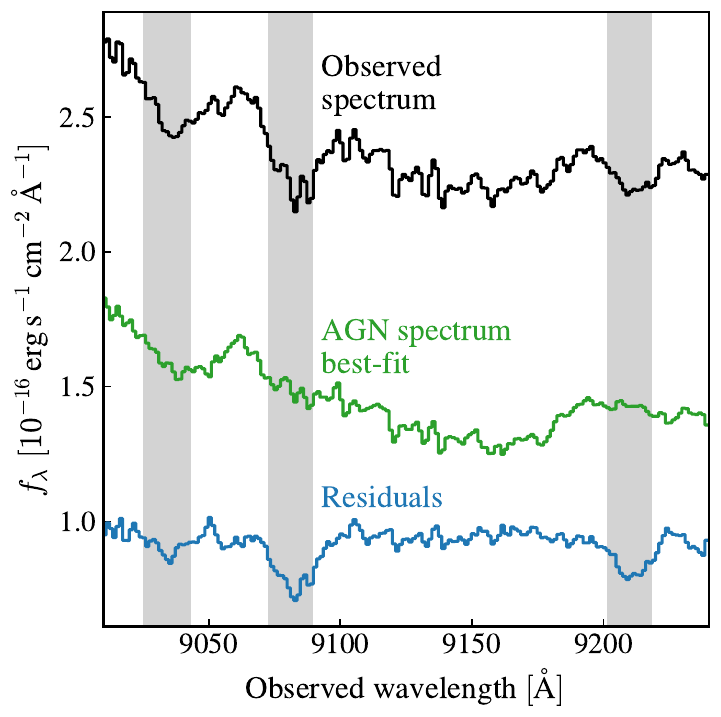}
\caption{\label{fig:agn_sub} Example of the AGN emission subtraction procedure. The shaded regions highlight the masked spectral windows encompassing the CaT. The AGN template has been vertically shifted to improve figure visualization. The residuals correspond to the observed spectrum minus AGN template.}
\end{figure}

For many of the high-luminosity AGNs the CaT lines close to the nucleus are significantly diluted by the underlying nonstellar continuum, to the extent that the stellar features are undetectable in the innermost spectrum extracted from the MUSE data cube. This allow us to consider this spectrum as an effective AGN template that can be used for modeling the AGN emission for the rest of the annuli-extracted spectra. Here, our main assumption is that the AGN is seen as a point source whose spectrum has been blurred by the MUSE PSF across the data cube. For those targets (see Table~\ref{tab:derived_quantities}), we add this empirical AGN template to the pool of spectra models, but as an additional component in {\tt pPXF} to avoid the AGN template broadening and shift in velocity space. For the stellar component we keep the {\tt pPXF} setup as detailed above. Thus, we model the AGN emission and the CaT simultaneously. We note that the additive and multiplicative polynomials affect all the templates when using {\tt pPXF} (see Equation~13 of \citealt{Cappellari2022} for more details). Figure~\ref{fig:agn_sub} shows an example for PG\,2130+099, where the AGN component is subtracted from the observed spectrum, and the residuals clearly show the CaT stellar features. 

\begin{figure*}
\includegraphics[width=2.0\columnwidth]{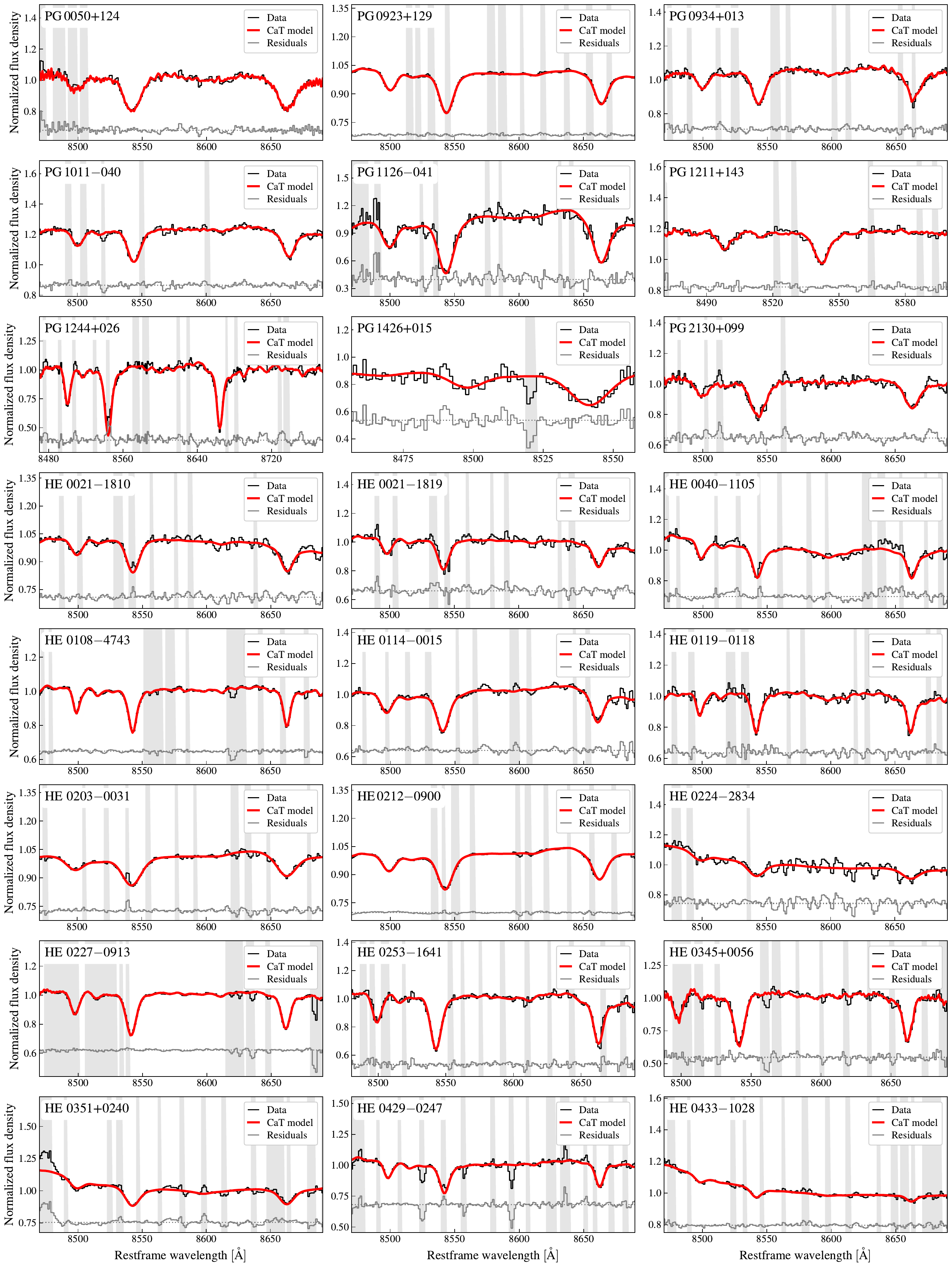}
\caption{\label{fig:CaT_fits_PG} Extracted spectrum at the CaT wavelength range for each host galaxy. For clarity, the model residuals have been shifted to the flux density level indicated by the dotted line. The gray shaded regions represent the spectral windows that are masked to avoid prominent sky-line features detected in the corresponding variance spectra.}
\end{figure*}

For each host galaxy, we select $\sigma_*$ from the model spectrum taken at an aperture radius as close as possible to the bulge half-light radius, following the host galaxy bulge properties derived from HST image modeling for the PG quasars \citep{Zhao2021} and MUSE white-light images for the CARS AGNs \citep{Husemann2022}. This radial constrain minimizes the aperture correction factor needed to convert $\sigma_*$ to $\sigma_e$ (see Section~\ref{sec:sigma_apcorr}). For our sources, the $R_e$ values are $\sim 1$\,kpc on average. For the selected spectra, we also compute the equivalent width (EW) and S/N of the Ca\,{\sc ii}\,$\lambda 8542$ absorption line. We choose Ca\,{\sc ii}\,$\lambda 8542$ as reference considering that the correct modeling of this stellar feature is enough to accurately quantify $\sigma_*$ from the CaT region \citep{Harris2012}. We present the selected spectra and their corresponding models in Figure~\ref{fig:CaT_fits_PG}. 

We use Monte Carlo resampling to derive the uncertainties in $\sigma_*$ \citep{Cappellari2022}. For each spectrum, the noise level (root-mean-square) is determined from the CaT model residuals. Then, we add simulated noise to the corresponding spectrum, assuming a normal distribution, and repeat the CaT fit. We iterate 1000 times to obtain a probability distribution for the best-fit parameters to estimate the $1\sigma$ uncertainties from the 16th and 84th percentiles. Table~\ref{tab:derived_quantities} provides the adopted aperture radius of the modeled spectrum and the $\sigma_*$ values for each host galaxy. We emphasize that this procedure also accounts for imperfect removal of the MUSE instrumental feature (Section~\ref{sec:sp_correction}), which is reflected in the spectrum model residuals (e.g., see HE\,0224$-$2834 and/or HE\,2128$-$0221 in Figure~\ref{fig:CaT_fits_PG}).

\begin{figure*}
\includegraphics[width=2.0\columnwidth]{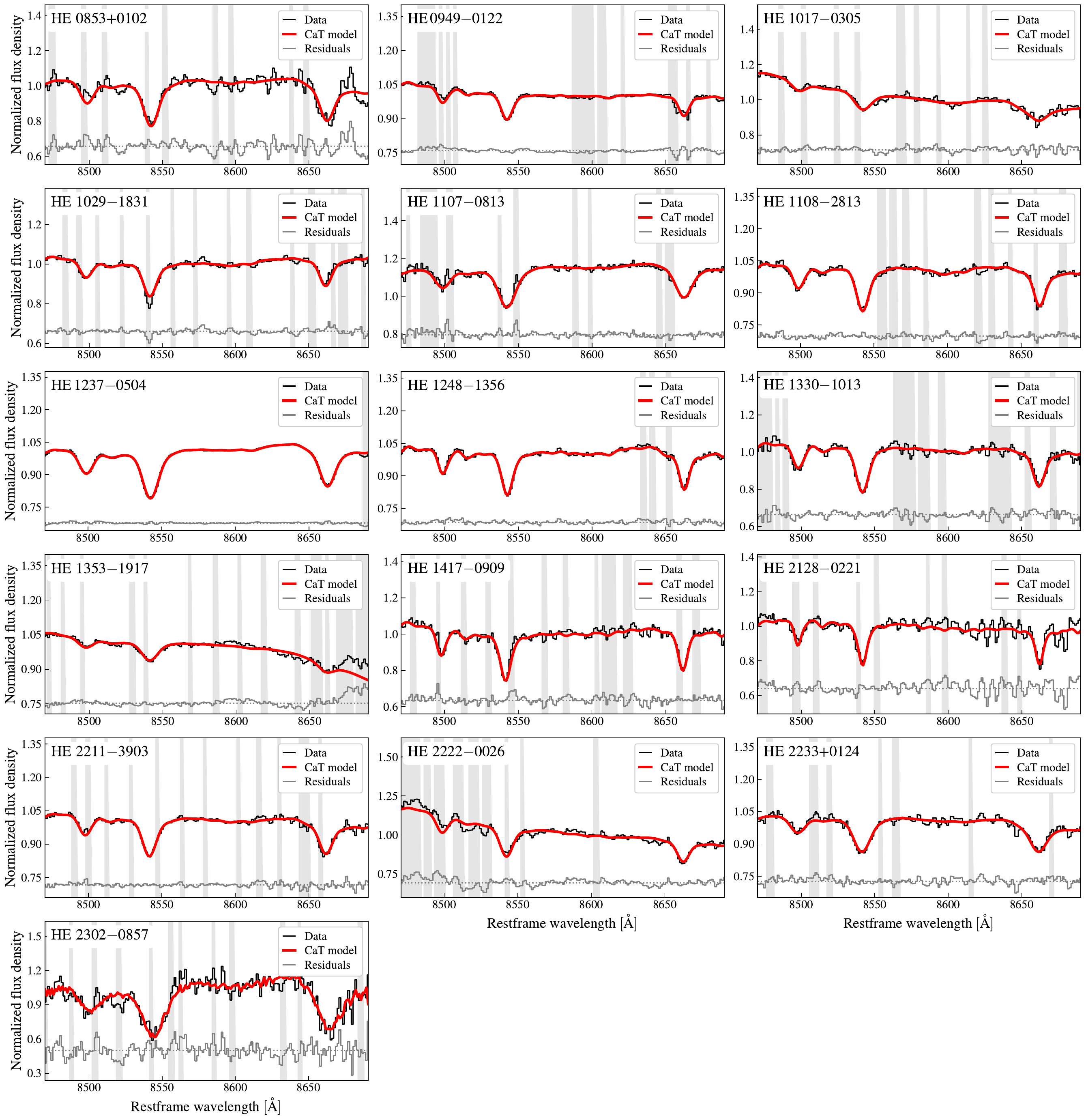}\\
{\textbf{Fig. 5.} continued.}
\end{figure*}

\subsection{Stellar velocity dispersion aperture correction}
\label{sec:sigma_apcorr}

For each AGN host galaxy, we use $\sigma_*$ to estimate $\sigma_e$. We follow the practice of the SAURON/ATLAS$^{\rm 3D}$ team for estimating $\sigma_e$, which derives the effective velocity dispersion from the luminosity weighted spectrum within $R_e$ (e.g., \citealt{Emsellem2007}). The option of using spatially resolved velocity fields to derive $\sigma_e$ (e.g., \citealt{Gultekin2009}) cannot be applied due to data limitation. However, both methods provide consistent $\sigma_e$ estimates for local massive systems with classical bulges (Figure 11 of \citealt{KormendyHo2013}). We provide further tests showing agreement between both procedures in Appendix~\ref{app:lw_sigmae}).

To estimate $\sigma_e$ from $\sigma_*$, we must apply an aperture correction that accounts for the radial gradient of $\sigma_*$ in galaxies \citep{Jorgensen1995,Cappellari2006}. We note that the aperture correction factors provided in the literature cannot be applied to our case because they were developed for spectra extracted using circular apertures rather than annular apertures. Here, we develop a correction recipe for estimating $\sigma_e$ from $\sigma_*$ when using annular apertures. By assuming that the bulge surface brightness profile can be well described by a S\'ersic model, an isotropic velocity dispersion (for simplicity), and a constant mass-to-light ratio, we find

\begin{equation}
\label{eq:ap_corr}
F_{\rm ap} =  \sum^2_{k=0} A_k(n) (R/R_e)^k,
\end{equation}

\noindent where $F_{\rm ap}$ corresponds to the correction factor by which $\sigma_*$ must be divided to obtain $\sigma_e$. The polynomials $A_k$ correspond to

\begin{equation}
\label{eq:ap_corr_coeff}
A_k(n) = A'_{k0} + A'_{k1} n^{-1} + A'_{k2}  n + A'_{k3} n^2, \\
\end{equation}

\noindent where the coefficients $A'_{kj}$ depend on the central bulge stellar mass profile, modeled by the S\'ersic index $n$. Equation~\ref{eq:ap_corr} is flexible enough to account for the blurring effects of the MUSE PSF, parameterized in terms of the ratio $\xi$ = PSF FWHM$/R_e$. We provide the coefficients $A'_{kj}$ in Table~\ref{tab:ap_corr_constants}. Equation~\ref{eq:ap_corr} is within $2\%$ for annular apertures located between 0.5 to $2.5\,R_e$, bulge S\'ersic indexes ranging from $n = 0.5$ to 8, and $\xi = 0 - 2.0$. This range of $\xi$ covers the typical properties of our MUSE observations (Table~\ref{tab:obssetup}). Appendix~\ref{app:ap_corr} gives more details about how we derived this numerical recipe. The $1\sigma$ error for the aperture correction factor can be estimated from the uncertainties of the bulge surface brightness S\'ersic profile parameters. However, these uncertainties are often underestimated due to systematics associated with non-axisymmetric components present in galaxies. Thus, we adopt a more conservative approach to compute the aperture correction factor uncertainty. For each target, we perform a Monte Carlo simulation where we vary $R_e$ and $n$ to compute $F_{\rm ap}$ using Eq.~\ref{eq:ap_corr}. The bulge half-light radius values are varied following a Gaussian distribution with width equal to the observation PSF FWHM (0\farcs2 for HST observations; \citealt{Zhao2021}). For the S\'ersic index, we assume values varying between 0.5 and 8 with no prior information. We iterate 1000 times to obtain a probability distribution for $F_{\rm ap}$ and compute the $1\sigma$ error from its standard deviation. We measure a typical aperture correction factor relative uncertainty of 11\% and 8\% for PG quasars and CARS AGNs, respectively.

\begin{figure}[!h]
\centering
\includegraphics[width=0.9\columnwidth]{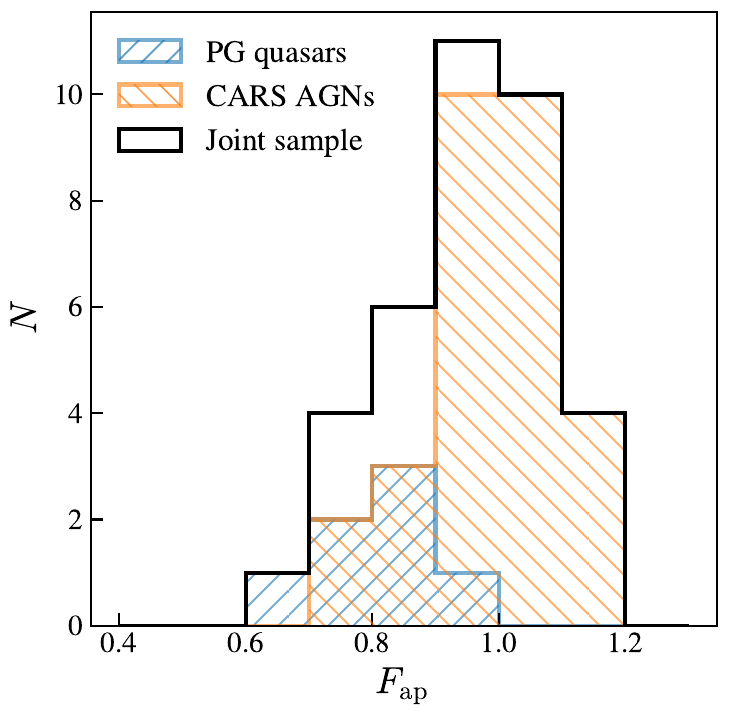}
\caption{\label{fig:aperture_corr} Annular aperture correction factor, $F_{\rm ap} \equiv  \sigma_* / \sigma_e $, for both AGN samples.}
\end{figure}

Figure~\ref{fig:aperture_corr} shows the correction factors applied to our sample, categorized into PG quasars and CARS AGNs. The annular aperture correction factors are $\sim 0.6-1.2$. On average, we find that the aperture correction factors are more significant for the PG quasar sample. This can be attributed to two factors: (1) the PG quasars tend to be at higher redshifts ($\langle z \rangle = 0.060$) compared with the CARS AGNs ($\langle z \rangle = 0.042$), resulting in spectra extracted farther away from the bulge $R_e$ due to their smaller projected sizes on the sky; and (2) the combination of the PSF blur and high AGN luminosity in the PG quasars limiting the extraction of $\sigma_*$ close to the bulge $R_e$. We note that the average observation seeing conditions for both AGN samples are similar ($\sim 0\farcs83 \pm 0 \farcs 26$).

\begin{table*}
        \small
        \def\arraystretch{1.2}
        \setlength\tabcolsep{2pt}
        \caption{\label{tab:ap_corr_constants} Coefficient values for computing the annuli aperture correction.}
        \vspace{0.2mm}
        \begin{tabular}{ccccccccccccc}
                \hline
                \hline
                $\xi$ & $A'_{00}$ & $A'_{01}$ & $A'_{02}$ & $A'_{03}$ & $A'_{10}$ & $A'_{11}$ & $A'_{12}$ & $A'_{13}$ & $A'_{20}$ & $A'_{21}$ & $A'_{22}$ & $A'_{23}$ \\
                \hline
                 0 & 1.160899 & 0.014853 & $-$0.002316 & $-$0.003392 & $-$0.125412 & $-$0.062056 & $-$0.082925 & 0.006842 & 0.000517 & $-$0.007138 & 0.023649 & $-$0.001860 \\
                 0.25 & 1.113832 & 0.028417 & 0.023769 & $-$0.004739 & $-$0.052492 & $-$0.077673 & $-$0.111226 & 0.008231 & $-$0.023532 & 0.000594 & 0.031723 & $-$0.002262 \\ 
                 0.50 & 1.027359 & 0.058682 & 0.060365 & $-$0.006613 & 0.070241 & $-$0.116072 & $-$0.145055 & 0.009891 & $-$0.060557 & 0.017315 & 0.040320 & $-$0.002689 \\ 
                 0.75 & 0.972627 & 0.077346 & 0.077192 & $-$0.007407 & 0.135668 & $-$0.136681 & $-$0.150332 & 0.010015 & $-$0.076557 & 0.026887 & 0.040020 & $-$0.002632 \\ 
                 1.00 & 0.943217 & 0.086742 & 0.082220 & $-$0.007629 & 0.161845 & $-$0.144046 & $-$0.142324 & 0.009502 & $-$0.079678 & 0.031006 & 0.036260 & $-$0.002408 \\ 
                 1.25 & 0.931808 & 0.089738 & 0.081267 & $-$0.007540 & 0.163137 & $-$0.142925 & $-$0.129668 & 0.008746 & $-$0.075490 & 0.031714 & 0.031682 & $-$0.002139 \\ 
                 1.50 & 0.928560 & 0.089947 & 0.078503 & $-$0.007362 & 0.155299 & $-$0.138913 & $-$0.117671 & 0.008037 & $-$0.069370 & 0.031187 & 0.027792 & $-$0.001909 \\ 
                 1.75 & 0.928987 & 0.089042 & 0.075475 & $-$0.007175 & 0.144786 & $-$0.134315 & $-$0.107702 & 0.007446 & $-$0.063277 & 0.030313 & 0.024825 & $-$0.001732 \\ 
                 2.00 & 0.930979 & 0.087749 & 0.072740 & $-$0.007011 & 0.134283 & $-$0.130023 & $-$0.099817 & 0.006978 & $-$0.057887 & 0.029448 & 0.022661 & $-$0.001603 \\ 
                 \hline
        \end{tabular}
        \tablefoot{The parameter $\xi$ refers to the ratio PSF FWHM$/R_e$.}
\end{table*}

\subsection{CaT EW and S/N-associated systematics}
\label{sec:Ion_source}

The scattered light from the nucleus in type~1 AGNs dilutes the host galaxy stellar spectrum by a featureless continuum. This spectrum dilution leads to a decrease in the EW and S/N for the stellar features (e.g., \citealt{Alexandroff2013}), introducing an additional source of uncertainty when deriving $\sigma_*$. Here, we investigate the effect of the AGN dilution for our measurements. 

\begin{figure*}[!h]
\centering
\includegraphics[width=1.8\columnwidth]{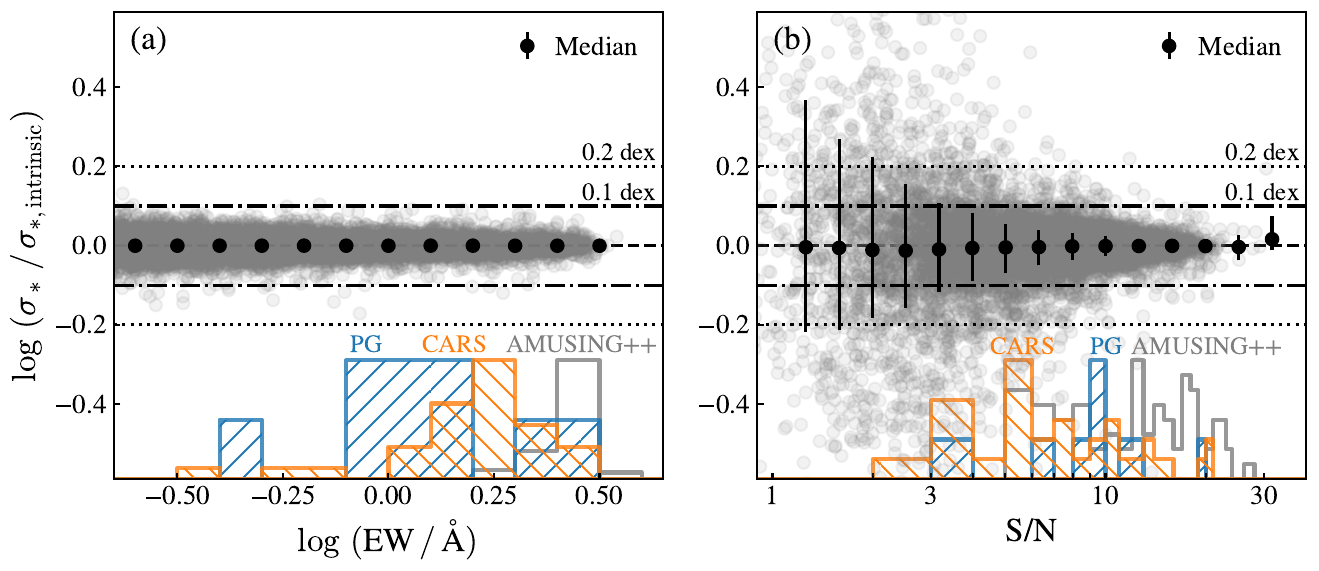}
\caption{\label{fig:simulation} Systematic uncertainty for the stellar velocity dispersion as a function of the (a) EW and (b) S/N of Ca\,{\sc ii}\,$\lambda8542$. The dashed line represents the equality between the observed $\sigma_*$ and the value obtained after degrading the spectra quality; dot-dashed and dotted lines represent the 0.1 and 0.2\,dex scatter, respectively. At the bottom of each panel, the normalized histograms show the distribution of EW or S/N values obtained for the PG quasars, CARS AGNs, and the inactive galaxy subsample taken from the AMUSING++ survey.}
\end{figure*}

We conduct a series of Monte Carlo simulations using MUSE observations of inactive galaxies taken from the AMUSING++ survey \citep{Lopez-Coba2020}. After removing Seyfert galaxies, mergers, low-quality data cubes (Ca\,{\sc ii}\,$\lambda8542$ line S/N\,<\,5), and matching the galaxies without AGN based on redshift ($0.015<z<0.09$), stellar mass ($10^{9.5}$--$10^{11.5}\,M_\odot$), and star formation rate (0.32--32\,$M_\odot\,$yr$^{-1}$), we use 75 MUSE observations of 9 E/S0, 14 spiral, 7 peculiar, and 45 unclassified galaxies. We extracted the spectra following the procedure outlined in Section~\ref{sec:sp_extract}, but applying circular apertures instead of annular apertures since we do not need to consider any nuclear emission. We mimic the effects of dilution by AGN emission on the line EW by adding a constant continuum and its associated Poisson noise until the recovered Ca\,{\sc ii}\,$\lambda8542$ EW reduces to $0.2\,$\r{A}. In a second test, we study the accuracy of the absorption-line width recovery against spectrum quality. This is done by adding Poisson noise to the inactive galaxy spectra until the resulting S/N of the Ca\,{\sc ii}\,$\lambda8542$ line is $\sim 1$. In both tests we encompass the Ca\,{\sc ii}\,$\lambda8542$ line EW and S/N values measured for the type~1 AGNs presented in this work.\footnote{Even though spectrum quality is commonly quantified by the S/N of the stellar continuum, we prefer to use the S/N of the Ca\,{\sc ii}\,$\lambda8542$ line as reference because AGN continuum subtraction is not needed for all sources (see Table~\ref{tab:derived_quantities}).} Figure~\ref{fig:simulation} summarizes the results from both simulations. Although the Ca\,{\sc ii}\,$\lambda8542$ line EW is significantly reduced owing to its dilution, the uncertainty of the $\sigma_*$ measurements remains controlled within $\sim$0.1\,dex for reasonable AGN-host galaxy contrast. This is because the added Poisson noise is not high enough to critically reduce the S/N of the absorption lines. We highlight this in our second test which emphasizes the stronger link between $\sigma_*$ recovery and the S/N of the stellar absorption feature. Our test results are in qualitative agreement with similar reports by \citet{Caglar2020}. Note that our analysis does not include the effect of CaT being influenced by blending with AGN emission lines, which are largely absent in this spectral region. The effect of CaT S/N on the uncertainty of $\sigma_*$ is already considered in our Monte Carlo resampling routine.

\subsection{BH mass}
\label{sec:BH_mass}

We estimate the BH masses following \citet{Ho2015}, who, building upon the work of \citet{Greene2005}, present recalibrated single-epoch virial mass estimators based on the updated virial coefficients for classical bulges and pseudo bulges \citep{Ho2014} and the BLR size and continuum luminosity relation of \citet{Bentz2013}. The BH masses (Table~\ref{tab:sample}) are computed as

\begin{equation}
\label{eq:Mbh_eq}
\log M_\bullet({\rm H\beta}) = \log \left[ \left( \frac{{\rm FWHM}({\rm H\beta})}{1000\,{\rm km\,s^{-1}}} \right)^2 \left( \frac{\lambda L_\lambda(5100\, \r{A})}{10^{44}\,{\rm erg\,s^{-1}}} \right)^{0.533} \right] + a,
\end{equation}

\noindent where $a = 7.03\pm0.02$ for classical bulges and $a = 6.62\pm0.04$ for pseudo bulges. The zero-point difference implies that host galaxies presenting pseudo bulges have 0.41\,dex lower BH mass for fixed H$\beta$ linewidth and AGN luminosity. The $M_\bullet$ uncertainties are conservatively estimated as the sum in quadrature of the scatter of the $M_\bullet-\sigma_e$ relation used to calibrate the BH mass prescription (0.29\,dex for classical bulges and ellipticals, and 0.46\,dex for pseudo bulges; \citealt{Ho2014}) and the scatter of the BLR-size relation (0.19\,dex; \citealt{Bentz2013}). Thus, we adopt $M_\bullet$ uncertainties of 0.35\,dex and 0.50\,dex for classical bulges and pseudo bulges, respectively. These estimates bracket the typical BH mass uncertainty of traditional single-epoch $M_\bullet$ prescriptions (e.g., $\sim 0.43\,$dex; \citealt{Vestergaard2006}). We classify the bulge type following \citet{Ho2015}, with classical bulges presenting $n > 2$ (but see \citealt{Gao2020}).

\section{Results}
\label{sec:Results}

In total, we successfully detect the CaT for 40 out of 42 host galaxies among the PG quasars and CARS AGNs. We model the CaT stellar features at a median distance of $\sim 1.1 \pm 0.5\,$kpc away from the host galaxy nucleus (Table~\ref{tab:derived_quantities}), corresponding to $\sim0.8$ times the bulge $R_e$. We consider the host galaxy morphology when constructing the annular apertures, minimizing the effect of host galaxy inclination when measuring $\sigma_*$. Figure~\ref{fig:CaT_fits_PG} provides a qualitative view for the spectral fits. Large residuals in the CaT models often correspond to sky lines that were masked during the fitting process (e.g., HE\,0429$-$0247), and/or inaccurate removal of MUSE instrumental features (e.g., HE\,0224$-$2834 and HE\,2128$-$0221). The latter are considered when estimating the uncertainties for the line widths. We detect the Ca\,{\sc ii}\,$\lambda8542$ line with high significance (S/N\,$\geq 10$) in 16 host galaxies, with low significance ($10>$\,S/N\,$\geq5$) in 18 cases, and with poor significance (S/N\,$<5$) in six host galaxies. We note that systems at higher redshifts tend to have lower S/N detection. We are unable to detect the CaT in two CARS AGNs because reliable spectra extraction for the host galaxies was not possible. These host galaxies are found too compact, with the AGN radiation dominating the observed emission. After discarding targets with S/N\,$<5$ detection, we obtain a total of 34 host galaxies with reliable $\sigma_e$ measurements.

\begin{figure}
\centering
\includegraphics[width=0.9\columnwidth]{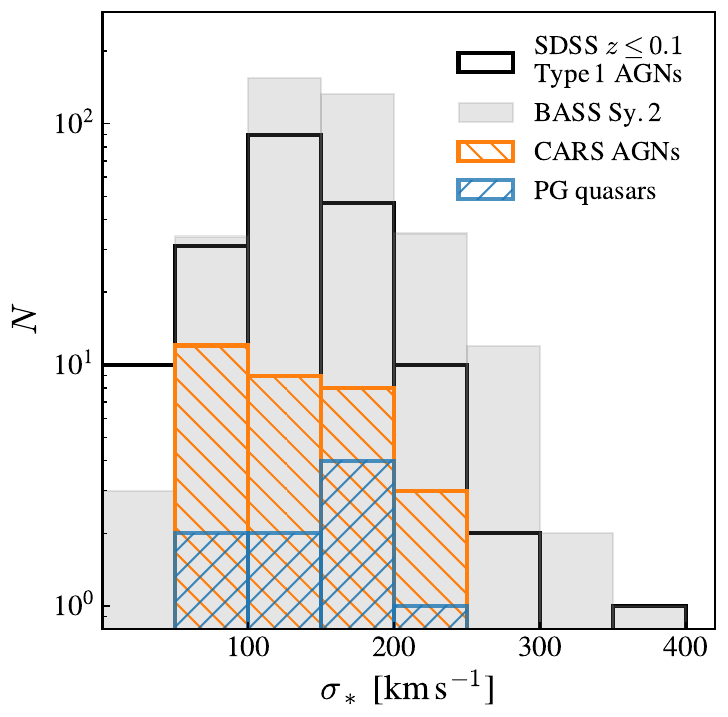}
\caption{\label{fig:sigma_star_dist} Distribution of the stellar velocity dispersion for the active galaxies. Literature estimates consider SDSS $z \leq 0.1$ type~1 AGNs \citep{Shen2008} and BASS type~2 Seyferts (including type~1.8 and 1.9 Seyferts; \citealt{Koss2022b}). We caution that the literature values were derived from spectra obtained using different aperture observation setups.}
\end{figure}

\subsection{Stellar velocity dispersion estimates}
\label{sec:sigma_star}

We provide the stellar velocity measurements in Table~\ref{tab:derived_quantities}.  We find that $\sigma_*$ values range from 60 to 230\,km\,s$^{-1}$, with median values of 152\,km\,s$^{-1}$ for PG quasars and 116\,km\,s$^{-1}$ for CARS AGNs. The typical $\sigma_*$ uncertainties are $\sim 10\% \pm 10\%$ ($\sim 14\,$km\,s$^{-1}$), although 8 sources have larger $1\sigma$ errors due to low-S/N detection of the CaT (e.g., HE\,0224$-$2834, HE\,1017$-$0305, HE\,1353$-$1917, PG\,1426+015). We note that the MUSE instrumental resolution at 9000\,\r{A} ($\sim 36\,$km\,s$^{-1}$; \citealt{Guerou2017}) is relatively low compared with the measured $\sigma_*$ values. Velocity dispersion values close to or below the instrumental resolution may have significant scatter or be overestimated \citep{Scott2018}. For instance, \citet{Koss2022b} exclude any $\sigma_*$ measurements within 20\% of the instrumental resolution limit for their sample of local type~2 quasars. In our case, all of our targets are above this limit, with HE\,2018$-$0221 presenting the lower $\sigma_*$ value, 67\% higher than the MUSE spectral resolution.

From the literature, we find few $\sigma_*$ measurements for our targets that can be used to make rough comparisons. Here, we report any $\sigma_*$ measurement irrespective of the stellar feature observed. \citet{Dasyra2007} presented near-IR $H-$band spectra modeling (mainly using CO stellar features; their Figure 2) for PG\,0050+124 ($\sigma_* = 188\pm36\,$km\,s$^{-1}$), PG\,1126$-$041 ($\sigma_* = 194\pm29\,$km\,s$^{-1}$), PG\,1426+015 ($\sigma_* = 185\pm67\,$km\,s$^{-1}$), and PG\,2130+099 ($\sigma_* = 156\pm18\,$km\,s$^{-1}$). Those spectra were obtained by using slits with widths of $\sim 1\farcs0-1\farcs4$, but avoiding the nuclear zones ($\lesssim 0\farcs3-0\farcs4$). \citet{Grier2013} presented adaptive optics-assisted Gemini near-IR IFU measurements, also taken in the $H$-band, for PG\,1426+015 ($\sigma_* = 211\pm15\,$km\,s$^{-1}$) and PG\,2130+099 ($\sigma_* = 167\pm19\,$km\,s$^{-1}$). Those measurements were extracted over annular apertures avoiding the AGN emission; they set annulus inner radius $\sim 0\farcs2-0\farcs4$ and outer radius $\sim 0\farcs6-1\farcs7$. \citet{Bennert2011} use Keck/LRIS long-slit spectroscopy to isolate the host galaxy emission and measure $\sigma_*$ for HE\,0203$-$0031 and HE\,0119$-$0118. Based on CaT modeling and apertures equal to the bulge $R_e$, they report $\sigma_* = 200\pm9\,$km\,s$^{-1}$ and $89\pm10\,$km\,s$^{-1}$, respectively. \citet{Busch2015} targeted HE\,1029$-$1931 with seeing-limited SINFONI observations in the $K$ band, reporting $\sigma_* = 104\pm20\,$km\,s$^{-1}$, using an aperture $\sim 0\farcs6$. Finally, for HE\,1237$-$0504, \citet{Caglar2020} estimate $\sigma_* = 145\pm4\,$km\,s$^{-1}$ by modeling the CO(2$-$0) absorption features observed with SINFONI (adaptive optics-assisted). Although different procedures and modeled stellar features make direct comparisons difficult, all of these literature measurements of $\sigma_*$ are consistent with ours (Table~\ref{tab:derived_quantities}). Using the adaptive optics observations as the main reference for duplicate literature values, we find a mean $\sigma_*$ ratio (literature divided by ours) of 1.04 with a standard deviation of 0.17.

Figure~\ref{fig:sigma_star_dist} shows the distribution for our $\sigma_*$ measurements. In contrast with the PG quasars, for the CARS sample we find more systems with $\sigma_* \lesssim 150 \,$km\,s$^{-1}$, which is consistent with the CARS survey covering less luminous AGNs with less massive BHs in less massive host galaxies and at lower redshifts (Section~\ref{sec:obs}). Additionally, we compare with the $\sigma_*$ distributions for two other AGN samples: local $z \leq 0.1$ type~1 AGNs \citep{Shen2008} from the SDSS and type~2 AGNs from the BAT AGN Spectroscopic Survey (BASS; \citealt{Koss2022}). We only select the BASS type~2 AGNs with CaT-based $\sigma_*$ values, corresponding to host galaxies at $z \lesssim 0.065$ \citep{Koss2022b}. Both samples provide a broader reference for type~1 AGNs, but it should be noted that their measurements may be less precise than ours for classical bulges and pseudo bulges because of aperture effects. For example, the SDSS data were obtained using $3\arcsec$-wide fibers. Despite this limitation, the BASS and SDSS samples still serve as a valuable comparison, especially considering the small correction needed for aperture effects. In the joint sample of PG quasar and CARS AGN host galaxies, we find a range of $\sigma_*$ values similar to those estimated for the SDSS $z \leq 0.1$ sample. The BASS type~2 AGNs tend to present a higher fraction of host galaxies with larger $\sigma_*$. This type~1-type~2 AGN dichotomy has been already noted by \citet{Koss2022b}, and can be linked to the differences in the properties of their host galaxies, with the type~2 AGN host galaxies likely being more massive (see also \citealt{Koss2011}).

After applying the aperture correction factors (Figure~\ref{fig:aperture_corr}), we report $\sigma_e$ values in the range of $60-250\,$km\,s$^{-1}$, with median values of $185\,$km\,s$^{-1}$ for the PG quasars and $108\,$km\,s$^{-1}$ for the CARS sample. The typical $\sigma_e$ uncertainties $\sim 16\% \pm 6\%$ ($\lesssim 25\,$km\,s$^{-1}$).

\subsection{The $M_\bullet-\sigma_e$ relation}
\label{sec:M_sigma}

\begin{figure}
\centering
\includegraphics[width=0.9\columnwidth]{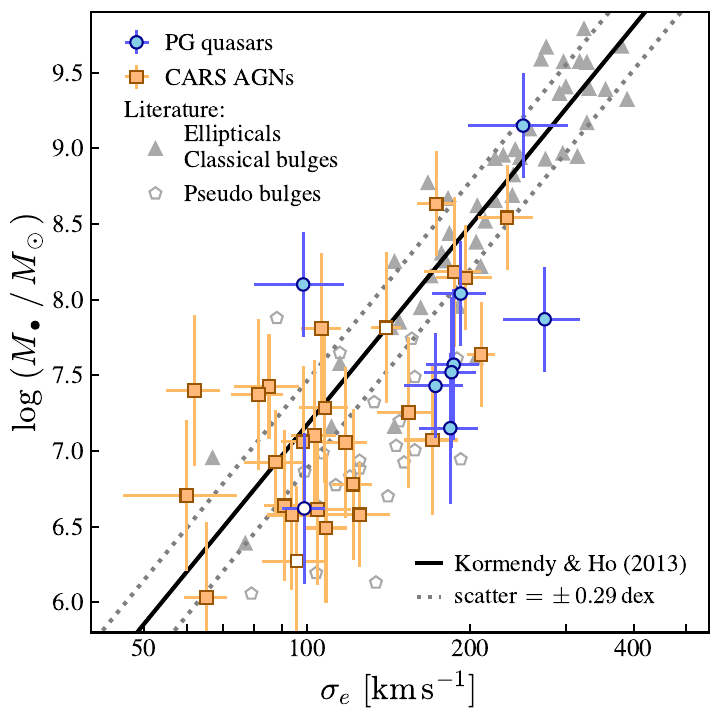}
\caption{\label{fig:mbh_sigma} $M_\bullet-\sigma_e$ relation for local type~1 AGNs and inactive galaxies. Colored open circles correspond to three AGNs for which the median sample aperture correction factor was adopted to estimate $\sigma_e$. The data for inactive galaxies are taken from \citet{KormendyHo2013}, as well as the best-fit relation (their Equation~7) and scatter.}
\end{figure}

Figure~\ref{fig:mbh_sigma} shows the relation between BH mass and $\sigma_e$ for the active galaxies. We primarily sample the parameter space at $M_\bullet \lesssim 10^8\,M_\odot$, where we detect a large scatter ($\sim$0.6\,dex) when considering the relation of \citet{KormendyHo2013} as reference. The CARS AGNs mainly span $M_\bullet-\sigma_e$ at $M_\bullet \lesssim 10^{7.5}\,M_\odot$, while the PG quasars complement CARS for higher BH masses. The lack of AGN data at higher BH masses is mainly due to the redshift upper limit for our sample ($z\lesssim0.1$). By performing a multivariate Cram\'er test \citep{Baringhaus2004}, using the \citet{KormendyHo2013} sample as reference for $M_\bullet \lesssim 10^8\,M_\odot$, we find no difference between the active and inactive galaxy samples ($p-$value = 0.69). The $p-$value value decreases to 0.42 when considering a BH mass upper limit of $10^{8.5}\,M_\odot$, but this is largely due to poor AGN sample statistics at $M_\bullet \gtrsim 10^8\,M_\odot$. 

\begin{figure}[!h]
\centering
\includegraphics[width=0.9\columnwidth]{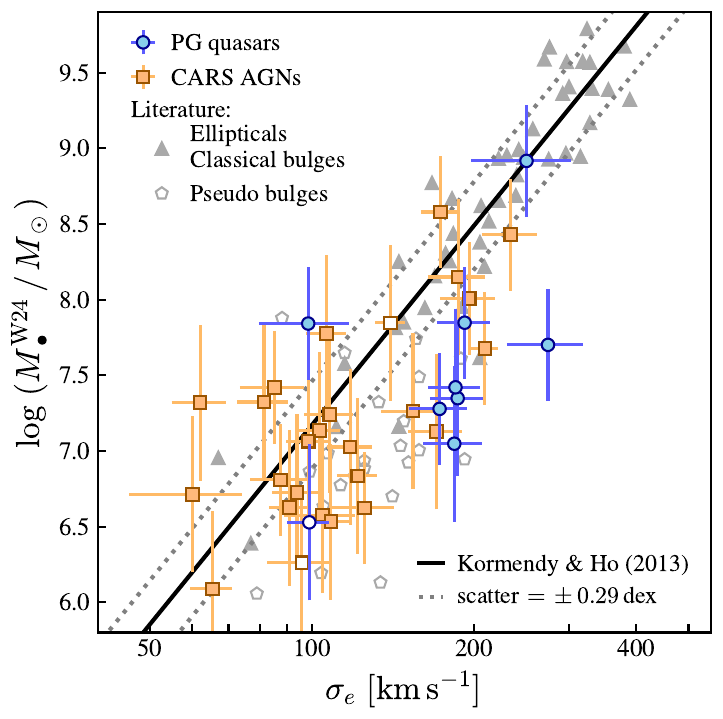}
\caption{\label{fig:mbh_sigma_W24}  $M_\bullet-\sigma_e$ relation for local type~1 AGNs, as in Figure~\ref{fig:mbh_sigma}, but updating the BH mass estimates following the BLR size-luminosity relation of \citet{Woo2024}.}
\end{figure}

Our BH mass estimates rely on the BLR size-luminosity relation of \citet{Bentz2013}. \citet{Woo2024} recently suggested a slightly shallower trend for this relation. They report typical BLR sizes being smaller for the more luminous AGNs  [$\lambda L_\lambda (5100\, \r{A}) \gtrsim 10^{44}$\,erg\,s$^{-1}$], suggesting less massive BHs because $M_\bullet$ scales linearly with BLR size. We explore this possibility by reestimating the BH masses following the BLR size-luminosity relation of \citet{Woo2024} and \citet{Ho2015}. For AGNs with $\lambda L_\lambda (5100\, \r{A}) \gtrsim 10^{44}$\,erg\,s$^{-1}$ (eight systems; six PG quasars) the BH masses decrease by $\gtrsim 0.15\,$dex. HE\,1248$-$1356 is the only system for which the BH mass increases by $\gtrsim 0.15\,$dex due to its low AGN luminosity [$\lambda L_\lambda (5100\, \r{A}) \lesssim 10^{42}$\,erg\,s$^{-1}$]. For all other AGNs, the BH masses vary by $\lesssim 0.15$\,dex, well below the typical $M_\bullet$ uncertainty range. Figure~\ref{fig:mbh_sigma_W24} presents our sample with updated BH masses ($M_\bullet^{W24}$). A multivariate Cram\'er test suggest no difference between the active and inactive galaxy samples ($p-$value = 0.49), in agreement with our previous report.

\begin{figure}
\centering
\includegraphics[width=0.9\columnwidth]{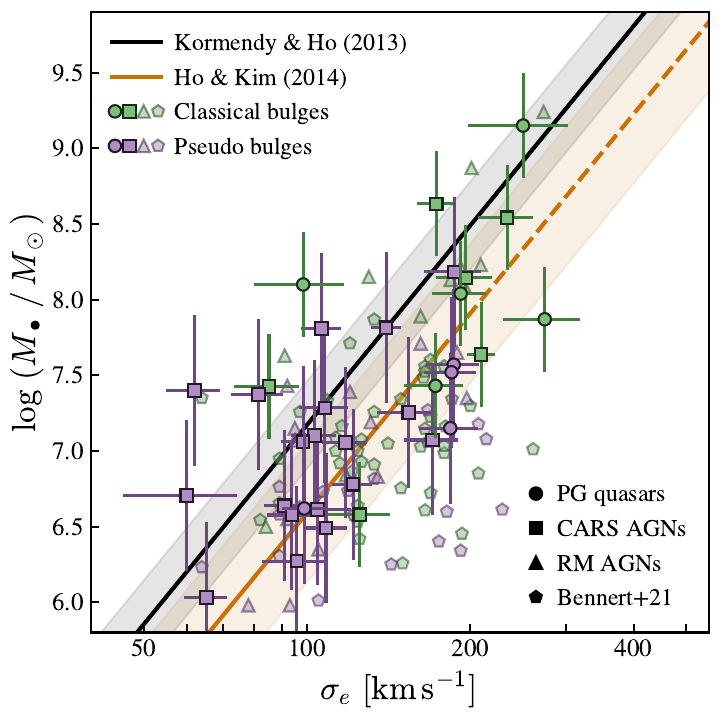}
\caption{\label{fig:mbh_sigma_onlyAGN}  $M_\bullet-\sigma_e$ relation for local type~1 AGNs, as in Figure~\ref{fig:mbh_sigma}, highlighting the best-fit relations for classical bulges \citep{KormendyHo2013} and pseudo bulges \citep{Ho2014}. We differentiate the systems by bulge type. The shaded regions represent the intrinsic scatter for both relations, 0.29\,dex and 0.46\,dex, respectively. The dashed line indicates the extrapolation of the best-fit relation for pseudo bulges at high BH masses. We also show the RM AGNs presented in \citet{Ho2014,Ho2015} and the local AGN sample of \citet{Bennert2021}.}
\end{figure}

\begin{figure}
\centering
\includegraphics[width=0.9\columnwidth]{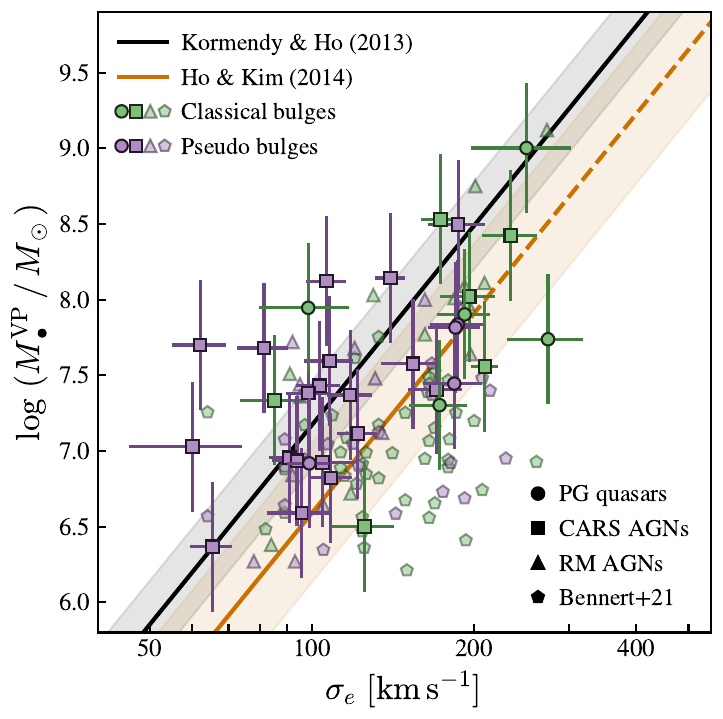}
\caption{\label{fig:mbh_sigma_VP06}  $M_\bullet-\sigma_e$ relation for local type~1 AGNs, as in Figure~\ref{fig:mbh_sigma_onlyAGN}, but adopting the single-epoch BH mass prescription of \citet{Vestergaard2006}. The $M_\bullet$ uncertainty is 0.43\,dex.}
\end{figure}

Considering the $1\sigma$ lower limit of the $M_\bullet-\sigma_e$ relation (0.29\,dex scatter) and the $1\sigma$ uncertainties of our $M_\bullet$ and $\sigma_e$ estimates,  Figure~\ref{fig:mbh_sigma} suggests four out of eight PG quasars and five out of 26 CARS AGNs lie below the $M_\bullet-\sigma_e$ relationship for ellipticals and classical bulges \citep{KormendyHo2013}, largely following the galaxies with pseudo bulges \citep{Saglia2016}. Finding some systems below $M_\bullet-\sigma_e$ might not be surprising, as we adopted the single-epoch BH mass prescription of \citet{Ho2015}, which systematically offsets the systems with pseudo bulge $-0.41\,$dex below the classical $M_\bullet-\sigma_e$ relation. To check whether this assumption drives our results, in Figure~\ref{fig:mbh_sigma_onlyAGN} we show $M_\bullet-\sigma_e$ differentiated by bulge type. For pseudo bulges, we adopt the best-fit reported by \citet[$\sim 0.46\,$dex scatter]{Ho2014}. They modeled the data collated in \citet{KormendyHo2013} keeping the slope fixed to that of the $M_\bullet-\sigma_e$ relation of classical bulges and ellipticals during the fitting process. We find that five out of nine of the AGN hosts located below $M_\bullet-\sigma_e$ for classical bulges and ellipticals are systems with pseudo bulges, suggesting that the adopted BH mass prescription may be effectively producing such trend. However, we report four AGN hosts with classical bulges below $M_\bullet-\sigma_e$ (e.g., HE\,0949$-$0122, PG\,2130+099). Literature data also show some systems with classical bulges following such trend. These AGN hosts are immune to the BH mass systematic described above. For example, by adopting the BH mass prescription of \citet{Vestergaard2006}, which does not differentiate AGN host galaxies by bulge type, to reestimate $M_\bullet$, all the systems with pseudo bulges shift upward in the $M_\bullet-\sigma_e$ plane (Figure~\ref{fig:mbh_sigma_VP06}), as expected. Only two AGN hosts with pseudo bulge remain below the $M_\bullet-\sigma_e$ relation for classical bulges and ellipticals after considering their BH mass uncertainty ($0.43\,$dex; \citealt{Vestergaard2006}). However, the four AGN host galaxies with classical bulges continue to line below the $M_\bullet-\sigma_e$ relation. We observe similar trends for the literature data. We note that a multivariate Cram\'er test indicates that our sample does not differ from that of inactive galaxies taken from \citet[p-value$\, < 0.01$]{KormendyHo2013} after adopting the BH mass prescription of \citet{Vestergaard2006}. 

Figure~\ref{fig:mbh_sigma} also shows two systems lying above the $M_\bullet-\sigma_e$ relation, HE\,0119$-$0118 and PG\,1211+143. When looking carefully, we find that HE\,0119$-$0118 has a barred host galaxy \citep{Husemann2022}, and PG\,1211+143 is compact \citep{Zhao2021}. It is well-known that measuring bulge properties in barred galaxies is difficult \citep[and references therein]{Gao2017}. We conjecture similar issues for PG\,1211+143, where the host galaxy morphology and AGN emission critically undermine accurate host galaxy image decomposition. Therefore, we cannot be certain of the accuracy of the offset with respect to $M_\bullet-\sigma_e$ for these two systems. It is possible that those measurements also reflect biased estimations of $\sigma_e$ due to poor host galaxy bulge characterization, in cases where the bulge is too compact and the region from where we extract $\sigma_*$ corresponds to that of a bar or disk. We caution that the results presented in this section depend, in part, on the adopted $M_\bullet$ scaling relationship (e.g., \citealt{Shankar2019}) and the criteria used to classify bulge type.

Another source of uncertainty arises from estimating $\sigma_e$ (and $\sigma_*$) itself. We have estimated $\sigma_e$ from a spectrum obtained by collapsing the host galaxy emission over an aperture. This method differs from that usually employed in observations that spatially resolve the galaxy kinematics (e.g., \citealt{Gultekin2009,KormendyHo2013}). Differential rotation, which is minimized by analyzing kinematic fields, may artificially broaden the spectrum obtained from an aperture, effectively increasing the estimated $\sigma_e$. While this effect may be minor for bulges, it may be significant for pseudo bulges due to their higher rotation support. \citet{Greene2005b} suggest that rotational broadening may be small. \citet{Bennert2011} find an average increase $\sim 10\%$ for $\sigma_e$ due to rotational broadening, but they note that rotational broadening could be more significant for edge-on systems (up to $\sim 40\%$, implying $\sim0.6$\,dex difference in BH mass offset). Note that, in the context of elliptical galaxies $\sigma_e$ is computed by co-adding spectra within apertures (e.g., \citealt{Cappellari2006}) even though they also typically rotate \citep{Cappellari2016}. We emphasize that we are comparing our sample with the inactive galaxy sample of \citet{KormendyHo2013}, whose $\sigma_e$ measurements were obtained from spatially resolved kinematics. \citet{KormendyHo2013} compared their estimates with aperture-based values provided by the SAURON/ATLAS$^{\rm 3D}$ team for massive galaxies with classical bulges, and found no major discrepancies (see their Figure 11b). We test the consistency of both procedures for systems with bulge and pseudo bulges by analyzing mock data in Appendix~\ref{app:lw_sigmae}. The mock data are build with galaxy properties encompassing those of our targets (Tables~\ref{tab:sample} and~\ref{tab:derived_quantities}). We find good agreement between both methods in estimating $\sigma_e$, with aperture-based values overestimating those obtained from spatially resolved kinematic maps by $\sim 2\% \pm 2\%$ for $\sigma_e \gtrsim 75\,$km\,s$^{-1}$. However, we caution that our test suggests a significant overestimation of $\sigma_e$ by the aperture-based method when the linewidths of the absorption features are comparable to the spectral resolution of the observations (see Appendix~\ref{app:lw_sigmae}, for more details).

One remaining uncertainty is whether to account for inclination effects when computing $\sigma_e$. \citet{Bennert2015} show that correcting $\sigma_e$ for inclination may increase the values by up to $\sim 40\%$. However, as \citet{Bennert2015} note, not correcting $\sigma_e$ for inclination is the common practice in the literature, meaning that any potential systematic bias would affect all galaxy samples, not just those hosting AGNs. Nevertheless, we find no correlation between BH mass offset from the $M_\bullet-\sigma_e$ relation and galaxy inclination for our sample.

\subsection{Deviations from the $M_\bullet-\sigma_e$ relation}
\label{sec:M_sigma_residuals}

Literature studies suggest that active galaxies with more efficiently accreting BHs tend to deviate from $M_\bullet-\sigma_e$, with the Eddington ratio\footnote{We compute the Eddington ratio as  $L_{\rm bol}/L_{\rm Edd}$, where $L_{\rm Edd} = 1.26 \times 10^{38}(M_\bullet/M_\odot)\,$erg\,s$^{-1}$.} being inversely correlated with the BH mass offset (e.g., \citealt{Shen2008,Ho2014}). The CARS AGNs and the PG quasars contain local AGNs with high $L_{\rm bol}/L_{\rm Edd}$, making our sample ideal for testing departures from the $M_\bullet-\sigma_e$ relation. The PG quasars mainly sample $L_{\rm bol}/L_{\rm Edd} \gtrsim 0.1$, while the CARS AGNs are spread all over the range $L_{\rm bol}/L_{\rm Edd} = 0.01-1$, although the majority of these systems present $L_{\rm bol}/L_{\rm Edd} \lesssim 0.1$. We compute the BH mass offset from $M_\bullet-\sigma_e$, using as reference Equation~7 of \citet{KormendyHo2013}:

\begin{equation}
\label{eq:delta_MBH}
\Delta \log\,(M_\bullet / M_\odot) = \log\left(\frac{M_\bullet}{M_\odot}\right) - 4.38\,\log\left(\frac{\sigma_e}{200\,{\rm km}\,{\rm s}^{-1}}\right) + 8.49.
\end{equation}

\begin{figure}
\centering
\includegraphics[width=0.9\columnwidth]{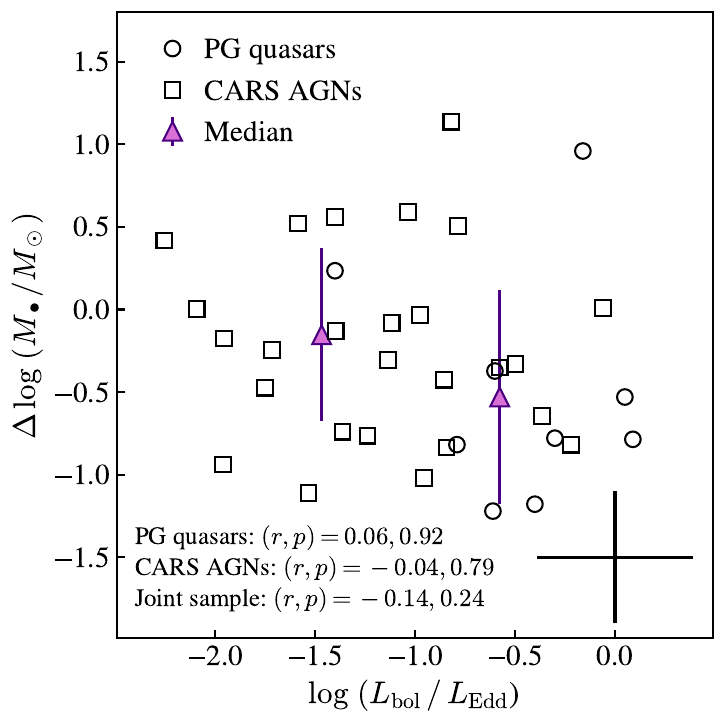} 
\caption{\label{fig:delta_MBH_vs_AGNprops} BH mass offset from the $M_\bullet-\sigma_e$ relation of \citet[their Equation~7]{KormendyHo2013} as a function of the Eddington ratio. The error bars in the bottom right corner represent the BH mass and Eddington ratio $1\sigma$ uncertainties. We detail the correlation coefficient $r$ and $p$-value for the PG quasars, CARS AGNs, and the joint sample.}
\end{figure}

\begin{figure*}
\centering
\includegraphics[width=0.9\columnwidth]{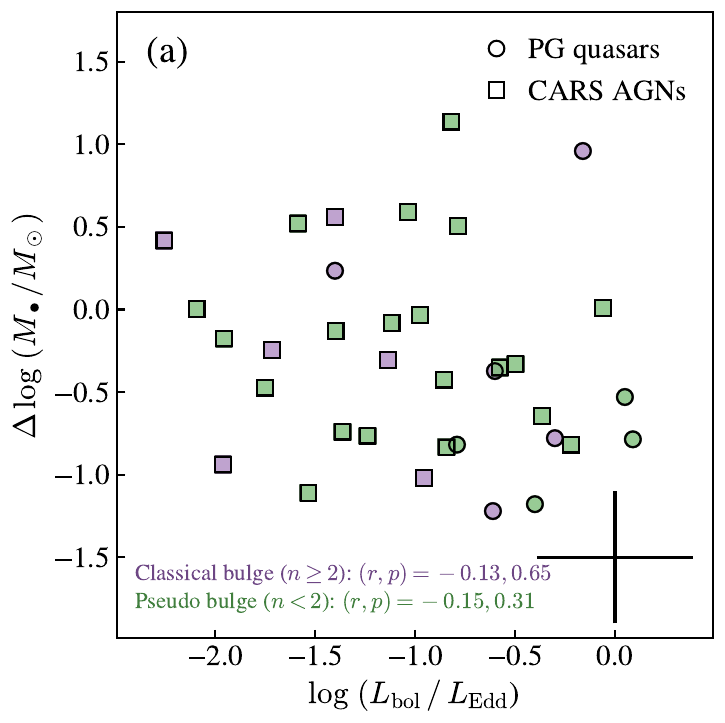}
\includegraphics[width=0.9\columnwidth]{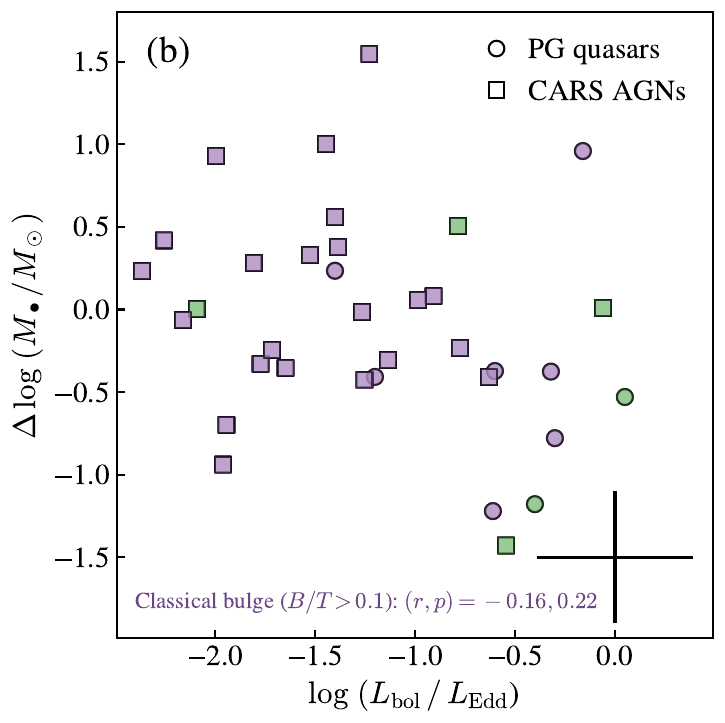}
\caption{\label{fig:delta_MBH_vs_AGNprops_colored} Similar to Figure~\ref{fig:delta_MBH_vs_AGNprops}, but color-coding the data by the host galaxy bulge type, which is differentiated by (a) S\'ersic index $n$ and (b) the $B/T$ ratio. We correct the BH mass-related quantities according to the bulge-type classification following the BH mass prescription of \cite{Ho2015}.}
\end{figure*}

Figure~\ref{fig:delta_MBH_vs_AGNprops} shows the BH mass offset as a function of Eddington ratio. We find no clear trends. Dividing the data into two bins of Eddington ratio, AGNs with $L_{\rm bol}/L_{\rm Edd} \leq 0.1$ present a median $M_\bullet$ offset of $-0.15$, with a scatter of 0.52\,dex. Objects with $L_{\rm bol}/L_{\rm Edd} > 0.1$ show a median BH mass departure equal to $-0.53$, with a scatter of 0.65\,dex. We compute Kendall $\tau$ correlation coefficients for both AGN samples and the joint sample (PG quasars plus CARS AGNs). In this analysis, we include the $1\sigma$ uncertainties of the quantities involved. We consider a correlation to be valid only if the probability of it occurring by chance is $\leq 0.05$. For the PG quasars, we compute $r = 0.06^{+0.22}_{-0.22}$ with a $p$-value of 0.92, while for the CARS AGNs we estimate $r = -0.04^{+0.09}_{-0.10}$ with a $p$-value of 0.79. For the joint sample, we compute $r = -0.14^{+0.08}_{-0.08}$ with a $p$-value of 0.24. No statistically significant correlation is seen. We find similar results when only considering host galaxies with CaT S/N\,$> 10$. Bearing in mind that galaxies with pseudo bulges are offset from $M_\bullet-\sigma_e$ and exhibit larger scatter \citep{KormendyHo2013}, we further distinguish host galaxies with classical or pseudo bulges. Figure~\ref{fig:delta_MBH_vs_AGNprops_colored} presents the data color-coded by bulge type, with the classical bulges labeled following two common classification schemes: the traditional S\'ersic index $> 2$ threshold \citep{Kormendy2004,Fisher2008}, and $B/T > 0.1$ \citep{Gao2020,Quilley2023}. For the first case, we find no correlation for either bulge type (classical bulges: $r=-0.13^{+0.18}_{-0.15}$, $p=0.65$; pseudo bulges: $r=-0.15^{+0.11}_{-0.11}$, $p=0.31$). For the second case, the limited statistics preclude us from computing a correlation coefficient for pseudo bulges. For galaxies hosting classical bulges, we find no significant correlation ($r=-0.16^{+0.10}_{-0.09}$, $p=0.22$) after correcting $L_{\rm Edd}$, $M_\bullet$, and BH mass offset considering our adopted bulge-type dependent BH mass prescription \citep{Ho2015}. However, we caution that using the $B/T$ ratio for classifying bulge type in AGN host galaxies is highly misleading, as it is known that the central spheroid can be overluminous from recent star formation activity \citep{Kim2019}. Furthermore, bulge model uncertainties are pernicious in the presence of a bright AGN glare. While we have some confidence in the AGN-host decomposition of the PG quasars, which were based on HST images \citep{Zhao2021}, the image analysis of the CARS AGNs was limited to MUSE white-light images \citep{Husemann2022}. We lack the statistics necessary to study subtle trends with respect to host galaxy morphology.

\begin{figure}
\centering
\includegraphics[width=0.9\columnwidth]{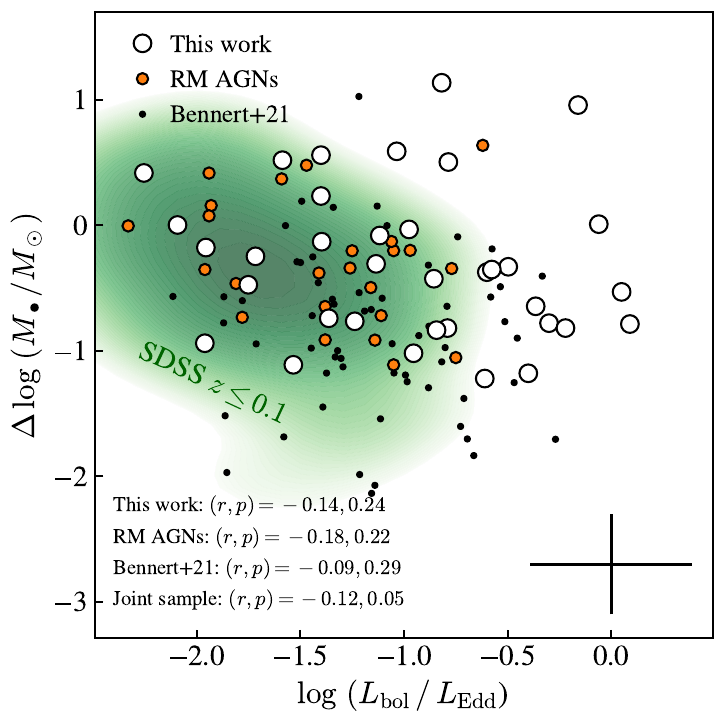} 
\caption{\label{fig:delta_MBH_vs_AGNprops_plus_literature} Similar to Figure~\ref{fig:delta_MBH_vs_AGNprops}, but we add the RM AGNs presented in \citet{Ho2014,Ho2015}, the local AGN sample of \citet{Bennert2021}, and the SDSS $z \leq 0.1$ type~1 AGNs provided by \citet{Shen2008}.}
\end{figure}

We complement our sample by including the less-luminous RM AGNs presented in \citet{Ho2014,Ho2015} and the local AGNs observed by \citet[Figure~\ref{fig:delta_MBH_vs_AGNprops_plus_literature}]{Bennert2021}. When combining the three data sets, the correlation coefficient slightly reduces to $r=-0.12^{+0.05}_{-0.05}$ with a $p$-value of 0.05. However, we note that the low $p-$value is mainly driven by the AGNs with Eddington ratio $< 0.01$ (only four objects). By excluding these AGN hosts we obtain $r = -0.09^{+0.04}_{-0.05}$ and $p=0.15$. The $p$-value is not low enough to imply a significant inverse correlation between BH mass offset and $L_{\rm bol}/L_{\rm Edd}$. 

\section{Discussion and summary}
\label{sec:Summ}

Early studies commonly reported that the AGN host galaxies tend to follow a different $M_\bullet-\sigma_e$ relation compared with that of inactive galaxies (e.g., \citealt{Greene2006b,Shen2008,Ho2014}). This was interpreted as a natural consequence of bright AGNs being preferentially hosted in late-type galaxies, considering that these systems tend to present pseudo bulges \citep{Kormendy2004} and pseudo bulge hosts lie below the $M_\bullet-\sigma_e$ relation of classical bulges and ellipticals \citep{Hu2008AGN,Greene2010,KormendyHo2013,Saglia2016,deNicola2019}. However, opposite reports have also been presented (e.g.,  \citealt{Nelson2004,Woo2010,Woo2013,Bennert2011,Bennert2015,Caglar2020}). By analyzing spatially resolved $\sigma_*$ profile data,  \citet{Bennert2021} suggest that AGN host galaxies with with a pseudo bulge follow the $M_\bullet-\sigma_e$ relation of classical bulges and ellipticals. Our measurements for the type~1 AGNs and quasars agree with the classical bulge--pseudo bulge dichotomy for $M_\bullet-\sigma_e$. Many of the PG quasars are located below the relation of \citet{KormendyHo2013} for classical bulges, in the regime of pseudo bulges, even though their central spheroids are classical bulges. Similar systems are also present in the sample of \citet[Fig~\ref{fig:mbh_sigma_onlyAGN}]{Bennert2021}. It is unlikely that underestimated BH masses is the sole cause in producing such trends. We have used the BH mass prescription of \cite{Ho2015}, which was explicitly calibrated for RM AGNs \citep{Ho2014} using the $M_\bullet-\sigma_e$ relation of inactive galaxies \citep{KormendyHo2013}. The difference in zero-point applied for host galaxies with classical bulges and pseudo bulges (Equation~\ref{eq:Mbh_eq}) can only account for a $M_\bullet$ offset up to 0.41\,dex for systems with a pseudo bulge. Still some AGN hosts with a pseudo bulge deviate $\gtrsim 1\,$dex from $M_\bullet-\sigma_e$ (Figure~\ref{fig:delta_MBH_vs_AGNprops_colored}a). However, we find average BH mass offsets from $M_\bullet-\sigma_e$ of $\sim -0.25 \pm 0.18$\,dex for AGN hosts with a classical bulge and $-0.31\pm 0.11$\,dex for those with a pseudo bulge, implying that any claim is on weak grounds. If we adopt the traditional single-epoch BH mass prescription of \citet{Vestergaard2006}, which does not distinguish by bulge type, many of the AGN hosts with a pseudo bulge shift upward, closer to $M_\bullet-\sigma_e$. Only a minor fraction ($16\%$) of AGN hosts remain below $M_\bullet-\sigma_e$ (Figure~\ref{fig:mbh_sigma_VP06}). However, in such a case, our AGN sample distribution on the $M_\bullet-\sigma_e$ plane would become significantly different from that of inactive galaxies taken from \citet[multivariate Cram\'er test $p$-value$\, <0.01$]{KormendyHo2013}. It is unclear why both samples may depart from each other, such that most of the disk-like AGN hosts may present overmassive BHs (or pseudo bulges with a lower $\sigma_e$) compared with the inactive spirals. BH masses could be underestimated if the BLR has a significant amount of dust, with $\sim 0.3-0.4\,$dex offsets reported for less luminous AGNs \citep{Caglar2020}. However, such BH mass offsets are not enough to explain our data, as mentioned above. Adopting the recent BLR-size luminosity relation of \citet{Woo2024} instead of that of \citet{Bentz2013} strengthens our finding, as \citet{Woo2024}'s BLR-size relation suggests less massive BHs for the more luminous AGNs. We note that the CaT is largely insensitive to the underlying stellar population properties \citep{Dressler1984}, suggesting that our findings should be robust against recent star formation activity that might induce systematic bias in AGN hosts. These findings are consistent with reports on active galaxies in the context of the BH mass-bulge mass relation, where some active galaxies have a systematically lower $M_\bullet$ than inactive galaxies for both cases when \citep{Molina2023} or when not differentiating \citep{Kim2008b,Zhao2021,Ding2022} by bulge type when estimating BH masses.
  
In our search for more subtle trends, we did not find any correlation between the Eddington ratio and offset from the $M_\bullet-\sigma_e$ relation, even after complementing our sample with the less-luminous RM AGN data. However, we cannot discount the possibility that the large uncertainties involved when estimating both the BH mass offset from $M_\bullet-\sigma_e$ and the Eddington ratio ($\sim 0.3-0.4\,$dex) may be washing out any potential weak correlation, further concealed by our small sample size. This is suggested in Figure~\ref{fig:delta_MBH_vs_AGNprops_plus_literature} by the SDSS type~1 AGN data, which can be interpreted as a rough reference for the population of type~1 AGNs covering a representative range of Eddington ratios at low redshifts \citep{Shen2008}.\footnote{We only consider the $\sigma_e$ estimates based on CaT modeling. The data were corrected by aperture effects; however, the bulge morphology was roughly approximated \citep[see][for more details]{Shen2008}.}$^,$\footnote{The BH masses and Eddington ratios of the SDSS sample were updated following \citet{Ho2015} for classical bulges.} The SDSS data show an inverse correlation between the $M_\bullet$ offset and $L_{\rm bol}/L_{\rm Edd}$ ($r=-0.31^{+0.05}_{-0.04}$, $p<0.01$). It is worth noting that sample selection effects are also concerning. On the one hand, as discussed in \citet{Shen2008}, an inverse correlation between $L_{\rm bol}/L_{\rm Edd}$ and the BH mass offset cannot be explained by the interdependence between $L_{\rm bol}$, $L_{\rm Edd}$, and $M_\bullet$, but it can be qualitatively produced if $\sigma_e$ and $\lambda L_\lambda (5100\, \r{A})$ are positively correlated. We ascertain that this is the case for the AGN sample analyzed here ($r = 0.25^{+0.05}_{-0.05}$, $p = 0.03$). On the other hand, applying a luminosity threshold for selecting active galaxies, as was done for the PG quasars \citep{Boroson1992} and CARS AGNs \citep{Schulze2009}, biases the samples toward AGNs with a high BH mass when considering the single-epoch virial mass estimate \citep{Shen2010}, making it harder to detect any residual correlation between the BH mass offset from $M_\bullet-\sigma_e$ and the Eddington ratio. The effects of these potential contending biases in producing the observed trend are, admittedly, uncertain. 

There have been reports that AGNs with high $L_{\rm bol}/L_{\rm Edd}$ being systematically below the BH mass--host stellar mass relation \citep{Volonteri2015,Shankar2019}. \citet{Zhuang2023} suggest that the level of BH accretion and star formation of an active galaxy is related to its position on the $M_\bullet-M_*$ plane, with BH growth outpacing the host galaxy stellar growth when the BH is undermassive. They show that most of the local AGN hosts below the $M_\bullet-M_*$ relation are late-type systems, with plenty of gas reservoirs as suggested by their blue color. We may be observing a similar trend in the $M_\bullet-\sigma_e$ plane as well. If the BHs are growing more rapidly than the host galaxy bulges in local AGNs and quasars with a high Eddington ratio, then they may culminate on the $M_\bullet-\sigma_e$ relation for classical bulges and ellipticals. \citet{Molina2023} note that in the more gas-rich PG quasars, the BHs can increase their mass significantly while the bulges may have already finished their mass buildup, unless the host galaxies undergo merging. The amount of mass that BHs can accrete may be rooted at the given formation stage of the host galaxy central stellar spheroid (e.g., \citealt{Merritt2004,Miralda2005,Angles2017a}), perhaps further modulated by a bursty nuclear stellar feedback episode \citep{Angles2017b}, or self-limited by feedback from the central engine surrounding the BH \citep{King2010,King2015}. \citet{Menci2023} suggest that the more efficiently accreting BHs tend to lie below the $M_\bullet-\sigma_e$ for classical bulges and ellipticals due to the large gas fuel supply available within the host galaxies and the relatively weak AGN feedback in the high Eddington ratio regime. We can neither rule out nor probe this possibility, but we note that \citet{Molina2023b} report a weak correlation between the Eddington ratio and molecular gas fraction for $z\lesssim0.5$ luminous AGNs, including the PG quasars and CARS AGNs. 
 
To summarize, this study used archival MUSE observations for 42 local ($z \lesssim 0.1$) type~1 AGNs and quasars taken from the CARS and PG quasar surveys. Our main goal was to measure the bulge stellar velocity dispersion from the CaT stellar features and investigate the location of the host galaxies in the $M_\bullet-\sigma_e$ relationship. We used annular apertures to extract the spectra and mitigate the effect of the AGN emission in diluting the host galaxy stellar features. Novel aperture corrections were developed to estimate accurate $\sigma_e$ in these systems. Our main results are as follows:

\begin{enumerate}

      \item The active galaxies have stellar velocity dispersion in the range $\sigma_* = 60-230$\,km\,s$^{-1}$. After correcting for aperture effects, bulge morphology, and observations' beam-smearing, we estimated $\sigma_e$ values in the range $60-250$\,km\,s$^{-1}$. 

      \item By assuming a BH mass prescription that differentiates between classical bulges and pseudo bulges, we find that the CARS AGNs and PG quasars span over the $M_\bullet$--$\sigma_e$ plane, with no statistical difference with respect to the inactive galaxies. The type~1 AGNs tend to preferentially follow the $M_\bullet$--$\sigma_e$ relation of elliptical and classical bulges, with $\sim 25\%$ AGN host galaxies located in the $M_\bullet$--$\sigma_e$ regime of pseudo bulges. We find two host galaxies above the local $M_\bullet$--$\sigma_e$ relation, but systematics associated with galaxy morphology are likely hindering their $\sigma_e$ estimates. If we do not differentiate by bulge type when estimating BH masses, the fraction of AGN hosts found below $M_\bullet$--$\sigma_e$ mildly reduces to $\sim 16\%$, and the sample distribution AGNs on the $M_\bullet$--$\sigma_e$ plane becomes inconsistent with that of the inactive galaxies.

      \item  We do not find any correlation between the BH mass offset from $M_\bullet$--$\sigma_e$ and the Eddington ratio, even after complementing our sample with the less luminous reverberation-mapped AGNs and AGN host galaxies with spatially resolved $\sigma_*$ profiles. However, we caution that this may be due to the large measurement uncertainties plus the small sample size.
      
\end{enumerate}

\begin{acknowledgements}
We thank the anonymous referee for constructive comments and suggestions. LCH was supported by the National Science Foundation of China (11721303, 11991052, 12011540375, 12233001), the National Key R\&D Program of China (2022YFF0503401), and the China Manned Space Project (CMS-CSST-2021-A04, CMS-CSST-2021-A06). K.~K.~acknowledges support from the Knut and Alice Wallenberg Foundation. This research has made use of the services of the ESO Science Archive Facility, and based on observations collected at the European Organization for Astronomical Research in the southern hemisphere under ESO programme IDs 094.B$-$0345(A), 095.B$-$0015(A), 097.B$-$0080(A), 099.B-0242(B), 099.B-0294(A), 0101.B$-$0368(B), 0103.B$-$0496(B) and 0104.B$-$0151(A), 106.21C7.002.

\end{acknowledgements}

\bibliographystyle{aa}
\bibliography{bibliography}

\begin{appendix}

\section{MUSE instrumental feature characterization}
\label{app:IF_charac}

\begin{figure*}[!htbp]
\centering
\includegraphics[width=0.9\columnwidth]{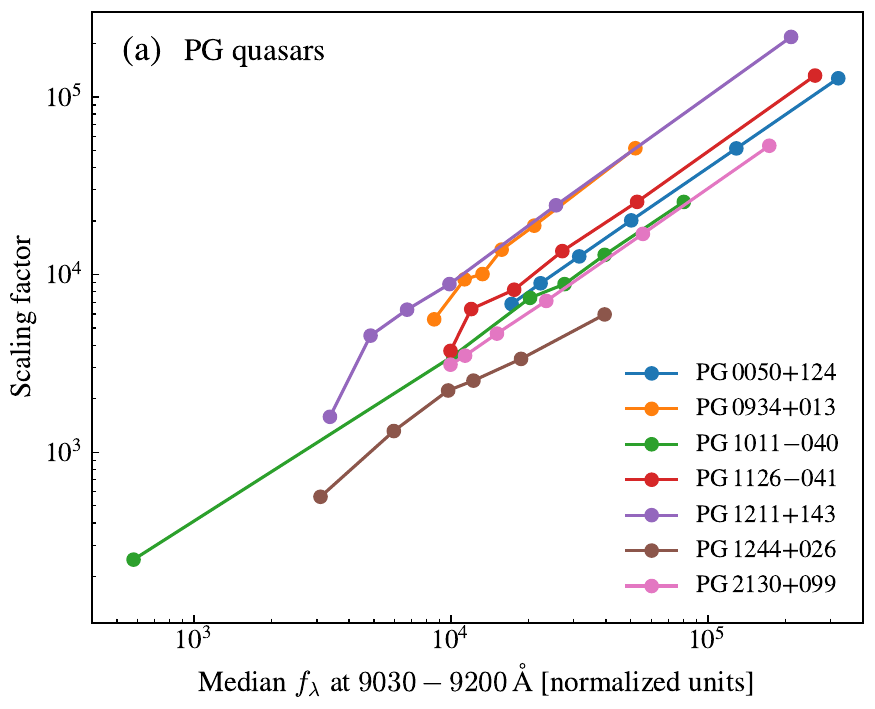}
\includegraphics[width=0.9\columnwidth]{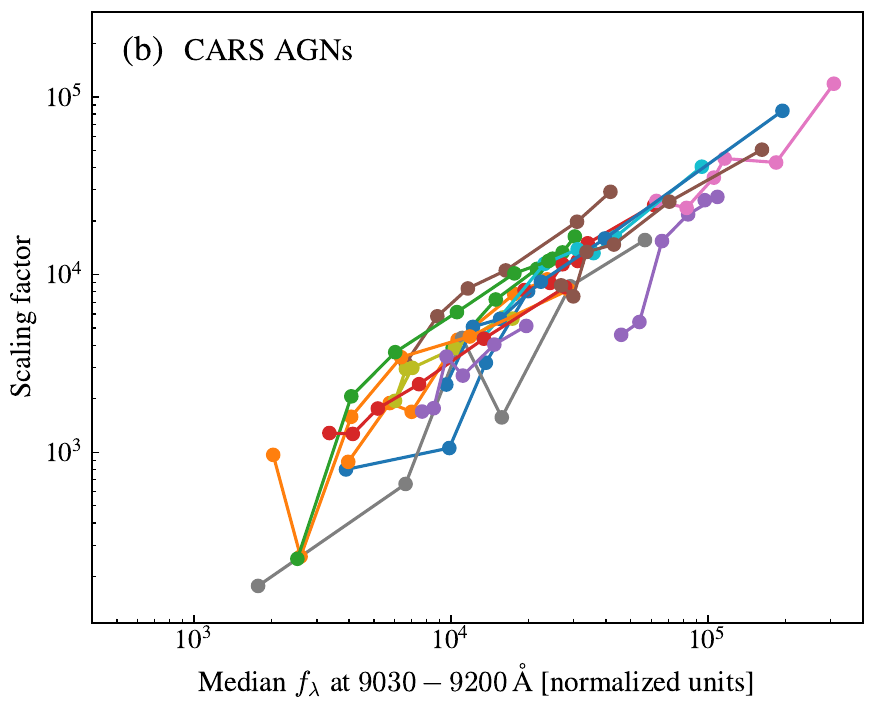}
\caption{\label{fig:if_charac_PG} Scaling factor for the instrumental feature template as a function of the annuli-extracted spectra median flux density over 9030--9200\,\r{A} (observer frame) for (a) PG quasars and (b) CARS AGNs. The solid lines connect the data points corresponding to each observation.}
\end{figure*}

For each target, we characterize and correct the MUSE data for an instrumental feature seen at 9060--9180\,\r{A} in the observer frame. We derive an instrumental feature template from field stars that are also observed by MUSE. For each target, the annuli-extracted spectra are modeled using an instrumental feature template multiplied by a scaling factor plus a power-law continuum component. A template flux normalization is determined by calculating the average of the ratio between the template scaling factor and local spectrum continuum level across all annular apertures. Figure~\ref{fig:if_charac_PG} shows the scaling factor as a function of the local continuum level for the PG quasar and CARS AGN samples. For all targets, the scaling factor is nearly proportional to the local continuum level, with minor variations corresponding to less accurate fits. By fitting a linear function to the data of each host galaxy, we recover slopes close to unity (as detailed for one case in Figure~\ref{fig:normalization_constant}), in agreement with our supposition that the observed spectrum features are related to instrument response to incoming flux. The zero point variation among the sources indicates that the flux density normalization of the instrumental feature varies for each observation.

\section{Luminosity weighted $\sigma_e$}
\label{app:lw_sigmae}

We estimate $\sigma_e$ from a spectrum obtained by spatially collapsing the data cube over a given aperture. That is, we estimate $\sigma_e$ from spatially unresolved data by adopting the convention of the SAURON/ATLAS$^{3D}$ team \citep{Emsellem2007}

\begin{equation}
G_e = \frac{\int_0^{R_e} I(R) G(R)\, R\,dR}{\int_0^{R_e} \, I(R)\, R\, dR},
\end{equation}
 
\noindent where $I(R)$ is the surface brightness and $G_e$ is the luminosity weighted mean of a quantity $G(R)$ within 1\,$R_e$. This approach differs from that commonly employed in $M_\bullet-\sigma_e$ relation studies, as these usually use spatially resolved observations to compute the luminosity weighted $\sigma_e$ as follows (e.g., \citealt{Gultekin2009})

\begin{equation}
\sigma_e^2 = \frac{\int_0^{R_e} (\sigma_V^2 + V^2) I(R)\,dR}{\int_0^{R_e} I(R)\,dR},
\end{equation}

\noindent where $V$ is the rotational component, and $\sigma_V$ is the velocity dispersion. \citet{KormendyHo2013} show that both procedures provide consistent $\sigma_e$ values for a local set of galaxies, independent of their bulge type (their Figure 11). In one of their tests, \citet{Bennert2015} show that their $\sigma_e$ measurements are consistent with SDSS-based estimates, with the later based on aperture-extracted spectra. 

\begin{figure*}[!h]
\centering
\includegraphics[width=2.0\columnwidth]{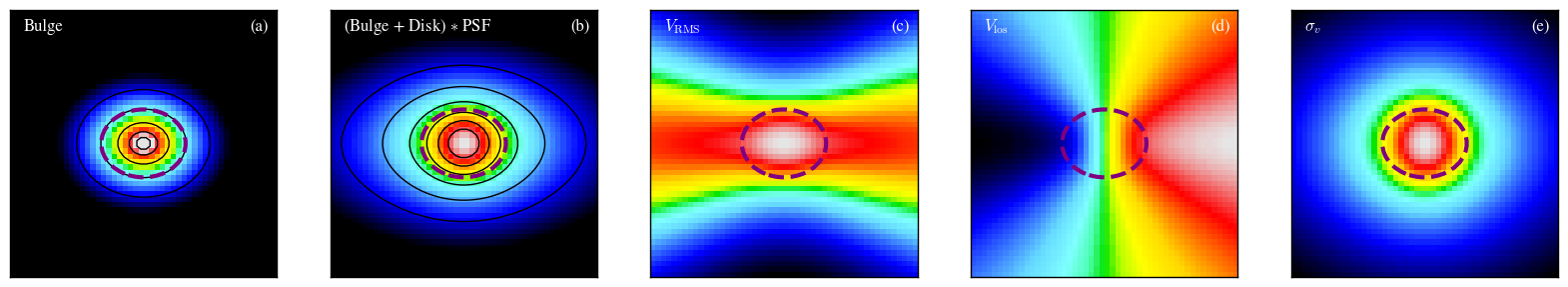}
\caption{\label{fig:galaxy_model_maps} Example of a galaxy model analyzed in our test. (a) Intrinsic surface brightness distribution of the bulge. (b) Galaxy bulge plus disk surface brightness distribution convolved by PSF. (b) ``RMS'' velocity $(V_{\rm RMS} \equiv \sqrt{\sigma_V^2 + V^2} \,)$ map. (c) Line-of-sight velocity field. (d) Velocity dispersion map. In all panels, the magenta dashed curve shows the bulge half-light radius. We apply the color scale in panels (a) and (b). The galaxy kinematics obtained from \textsc{jampy} \citep{Cappellari2008}.}
\end{figure*}

\begin{figure}[!hb]
\centering
\includegraphics[width=0.9\columnwidth]{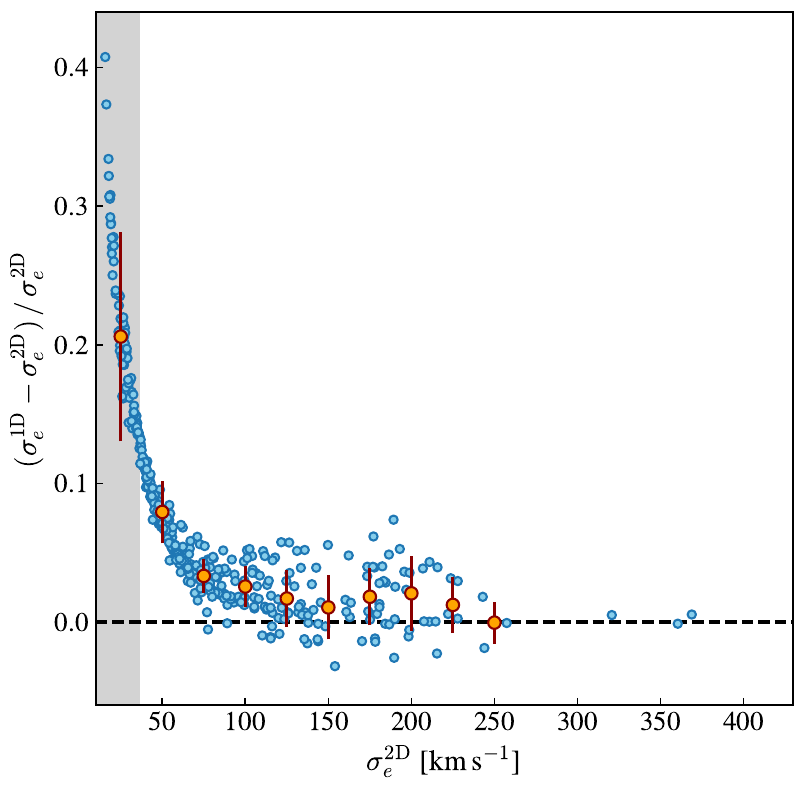}
\caption{\label{fig:1D_vs_2D_sigmae} Relative difference between spatially resolved and aperture-based $\sigma_e$ measurements. The orange circles represent the mean trend over 25\,km\,s$^{-1}$-wide bins. The error bars correspond to the data scatter. The shaded region represents the spectrum resolution of MUSE in terms of line width ($\sim$\,FWHM/2.355).}
\end{figure}

Here, we test if both procedures used to estimate $\sigma_e$ provide consistent results for mock data with properties similar to our sample. We model a galaxy as disk plus bulge, with each surface brightness component described by a S\'ersic profile when projected on the sky. After setting the galaxy profile, we compute the second velocity moment map using \textsc{jampy} \citep[Figure~\ref{fig:galaxy_model_maps}]{Cappellari2008}. We consider a PSF FWHM $= 1 ''$. From this map, we estimate effective velocity dispersion within the bulge $R_e$ following the common practice ($\sigma_e^{\rm 2D}$; \citealt{Gultekin2009}). To build a three-dimensional data cube, we adopt a K-star spectrum template taken from the INDO-U.S. stellar spectral library of \citet{Valdes2004}, and the two-dimensional surface brightness and kinematic maps for our galaxy model. The kinematic maps set the spectra Doppler shift and broadening at each data cube pixel that we must apply for. We convolve the spectra by the MUSE line spread function, accounting for the stellar template spectral resolution (${\rm FWHM} = 1.35\,$\r{A}; \citealt{Beifiori2011}). Finally, we collapse the data cube over a circular aperture equal to the bulge $R_e$ to extract the spectrum and obtain the effective velocity dispersion ($\sigma_e^{\rm 1D}$). We repeat this process 500 times, varying the galaxy parameters randomly following uniform distributions with parameter ranges that encompass the observed values for our sample (Table~\ref{tab:sample} and ~\ref{tab:derived_quantities}). Specifically, we consider galaxies with total mass between 9.6--11.2\,$M_\odot$ and bulge-to-total ratio in the 0.01--0.97 range. S\'ersic index values in the 0.5--8 range for the bulge and 0.5--3 range for the disk. We set bulge effective radius in the range of 0$\farcs$5--5$''$, and the disk size can be 1.1--5 times larger. The bulge and disk can have different observed inclination on the sky. For both cases, the minor-to-major axis ratio range is 0.3--0.95, that is inclination angles between $0^\circ$ (face-on) and $88^\circ$ (edge-on). The axis ratio lower limit is set considering an intrinsic component axis-ratio of 0.2. Figure~\ref{fig:1D_vs_2D_sigmae} shows our test results. We find a good agreement between both estimates; the mean trend shows that $\sigma_e^{\rm 1D}$ agrees with $\sigma_e^{\rm 2D}$ by $\lesssim 10\%$ for $\sigma_e^{\rm 2D} \gtrsim 50$\,km\,s$^{-1}$. For $\sigma_e^{\rm 2D} \gtrsim 75$\,km\,s$^{-1}$,  $\sigma_e^{\rm 1D}$ overestimates $\sigma_e^{\rm 2D}$ by merely $\sim 2\%\pm2\%$. Such factor is negligible considering the typical uncertainty of our measurements ($\sim 10\%$; Section~\ref{sec:sigma_star}). For $\sigma_e^{\rm 2D} \lesssim 50$\,km\,s$^{-1}$, we find that $\sigma_e^{\rm 1D}$ increasingly overestimates $\sigma_e^{\rm 2D}$ for lower values. However, both quantities remain closely correlated, with no significant dependence on the properties of the simulated systems. The overestimation of $\sigma_e^{\rm 2D}$ by $\sigma_e^{\rm 1D}$ for $\sigma_e^{\rm 2D} \lesssim 50$\,km\,s$^{-1}$ seems to be related to measuring velocity dispersion values close to or below the spectral resolution of the data \citep{Scott2018}, which is $\sim 36\,$km\,s$^{-1}$ for MUSE at $9000\,\mathrm{\AA}$ \citep{Guerou2017} and our mock data. Given the measured $\sigma_*$ values for our sample (Table~\ref{tab:derived_quantities}), this analysis suggests that we can safely compare our estimates with the spatially resolved $\sigma_*$ measurements provided in the literature.

\section{Annular aperture correction factor}
\label{app:ap_corr}

Aperture correction factors must be applied for estimating $\sigma_e$ from projected (line-of-sight) second velocity moment ($\sigma_{\rm los}$) measurements. We derive aperture correction factors under the assumption that the host galaxy bulge surface brightness emission on the sky is well-described by a \cite{Sersic1963} radial profile

\begin{equation}
\label{eq:sersic_prof}
I(R) = I_0 \exp(-b_n (R/R_e)^{1/n}),
\end{equation}

\noindent where $I_0$ is the central intensity and $b_n$ is the constant that sets $R_e$ as the half-light radius. By further assuming an isotropic velocity dispersion, $\sigma_{\rm los}$ can be calculated as \citep{Baes2019}

\begin{equation}
\label{eq:projsigma_prof}
\sigma_{\rm los}^2(R) = \frac{2}{I(R)} \int_R^\infty \frac{\nu(r) M(r) \sqrt{r^2-R^2} dr}{r^2},
\end{equation}

\noindent where $\nu(r)$ and $M(r)$ are the three-dimensional luminosity density and mass profiles, respectively. The $\nu(r)$ and $M(r)$ profiles are given by \citep{Binney2008}

\begin{equation}
\begin{array}{l}

\nu(r) = \frac{-1}{\pi} \int_R^\infty \frac{{\rm d}I(R)}{{\rm d}R} \frac{dR}{\sqrt{r^2-R^2}}, \\
M(r) = 4 \pi \left( \frac{M}{L} \right) \int_0^r \nu(r')r'^2 dr',

\end{array}
\end{equation}

\noindent where $M/L$ is the mass-to-light ratio, assumed to be constant in this work. The $\nu(r)$ and $M(r)$ profiles can be computed in terms of the Fox $H$ function \citep{Mathai2009} following Equations~(17) and (25) of \citet[see also \citealt{Baes2011}]{Baes2019}. For completeness, we also provide the formula for computing the projected (luminosity weighted) second velocity moment when adopting a circular aperture:

\begin{equation}
\label{eq:circap_projsigma_prof}
\sigma_{\rm los}^2(<R) = \frac{\int_0^R I(R') \sigma_{\rm los}^2(R') R' dR'}{\int_0^R I(R') R' dR'}.
\end{equation}

The PSF effect is considered following \citet[see also \citealt{Cappellari2008}]{Emsellem1994}. We compute the PSF-convolved line-of-sight second velocity moment ($\widetilde{\sigma}_{\rm los}$) as

\begin{equation}
\label{eq:convsigma_prof}
\widetilde{\sigma}_{\rm los}^2(R) = \frac{(I \sigma_{\rm los}^2)(R)\ast {\rm PSF}(R)}{I(R) \ast {\rm PSF}(R)},
\end{equation}

\noindent where the PSF follows a \cite{Moffat1969} profile with $\beta \approx 2.2$ for MUSE data \citep{Bacon2017,Guerou2017}.

\begin{figure*}[!h]
\centering
\includegraphics[width=1.8\columnwidth]{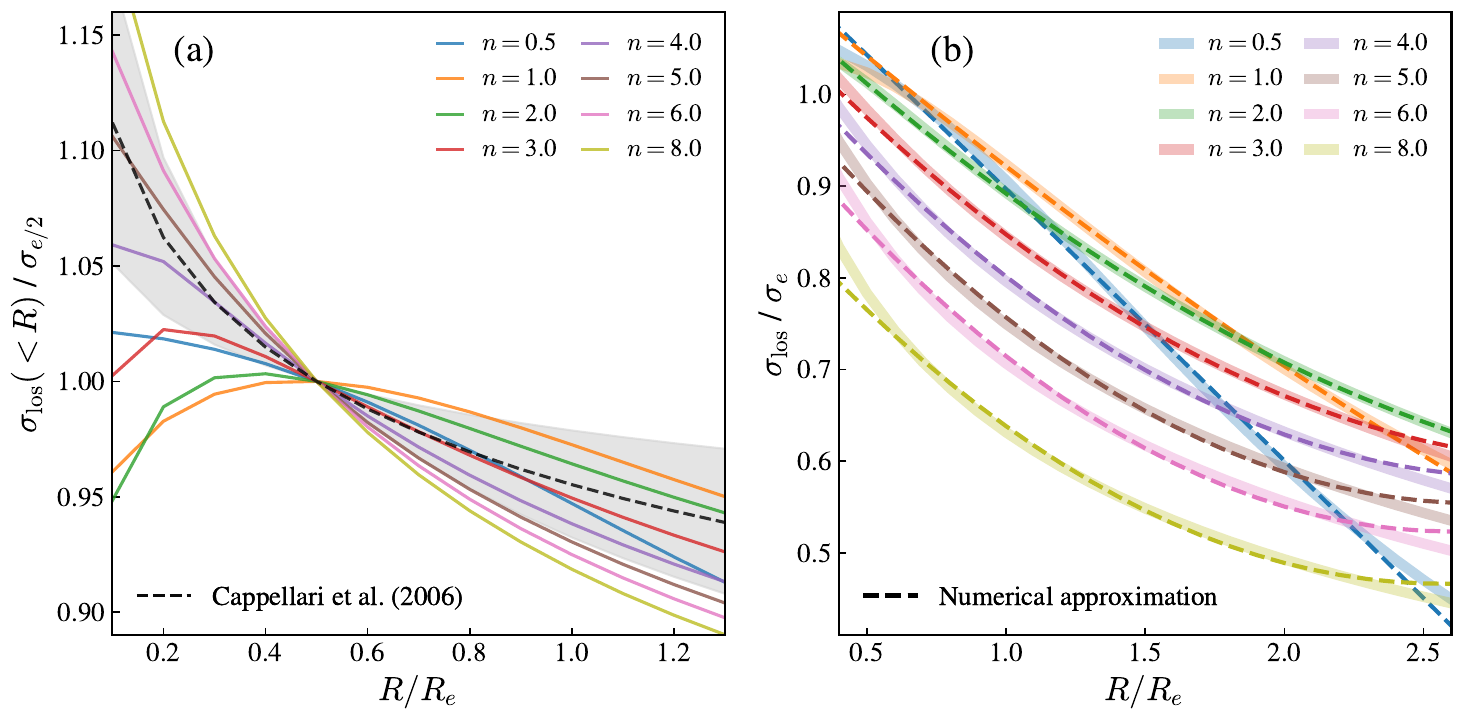}
\caption{\label{fig:ap_corr_intr} Aperture correction as a function of (a) the circular aperture size and (b) annular aperture location. The circular aperture correction values are normalized to the effective velocity dispersion at $R_e/2$ ($\sigma_{e/2}$), following \citet{Cappellari2006}. We also highlight the empirical mean circular aperture correction provided by \citet{Cappellari2006} for local E/S0 galaxies, with the shaded region representing its scatter. The numerical approximation for the annular aperture correction factors corresponds to that of Equation~\ref{eq:ap_corr}, using the input coefficients presented in Table~\ref{tab:ap_corr_constants}.}
\end{figure*}

Figure~\ref{fig:ap_corr_intr} shows the annular and circular aperture correction factors as a function of radius, normalized to $R_e$, and for different values of the S\'ersic index $n$ in the case of no PSF convolution. When using circular apertures, the correction factor is small ($\sim 10\%$), but non-negligible. The analytic formula accurately reproduces the empirical correction found by \citet{Cappellari2006} for local E/S0 galaxies. When considering annular apertures, the second velocity moment correction factor can be as high as 50\%, depending on the S\'ersic index and/or reasonable values for annular distance from $R_e$. Observing bulges with large S\'ersic index (e.g., classical bulges) generally requires applying more significant corrections for $\sigma_{\rm los}$ to estimate $\sigma_e$ at a fixed radius. Note that at large annular radii additional systematics could influence the estimation of $\sigma_e$ (e.g., the kinematics of the disk). Figure~\ref{fig:ap_corr_intr} highlights the difference between adopting annular and circular apertures for estimating $\sigma_e$ and the importance of measuring $\sigma_{\rm los}$ as close as possible to $R_e$. 

\begin{figure*}[!h]
\centering
\includegraphics[width=1.8\columnwidth]{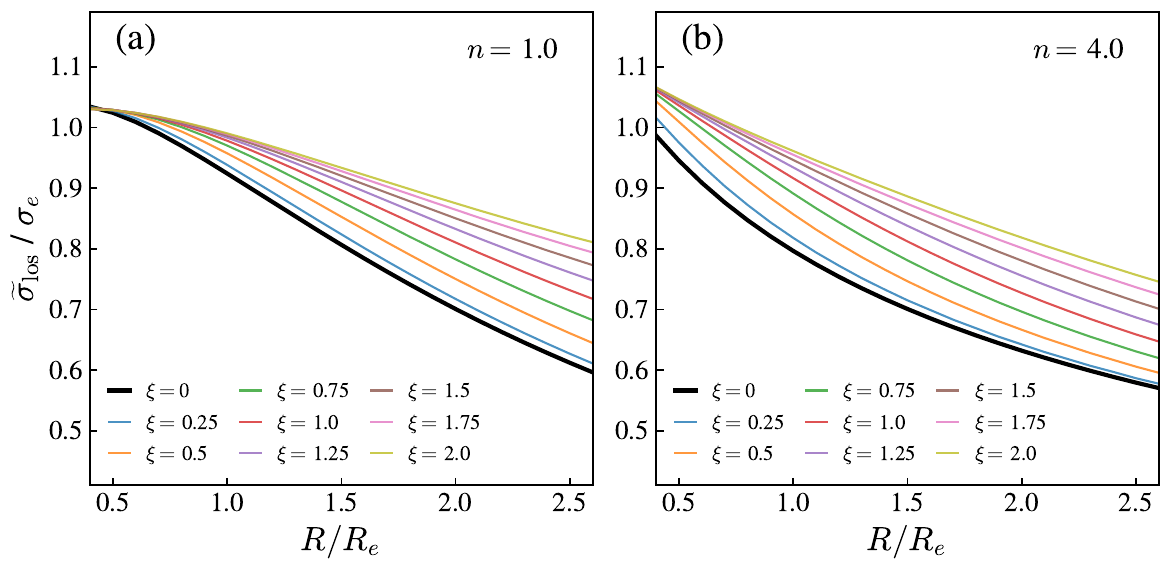}
\caption{\label{fig:ap_conv_effect} PSF convolution effect on the annular aperture correction factor for different values of PSF FWHM, normalized to S\'ersic model $R_e$. We only show results for two S\'ersic surface brightness profile cases (a) $n=1$ and (b) $n=4$.}
\end{figure*}

\begin{figure*}[!h]
\centering
\includegraphics[width=1.8\columnwidth]{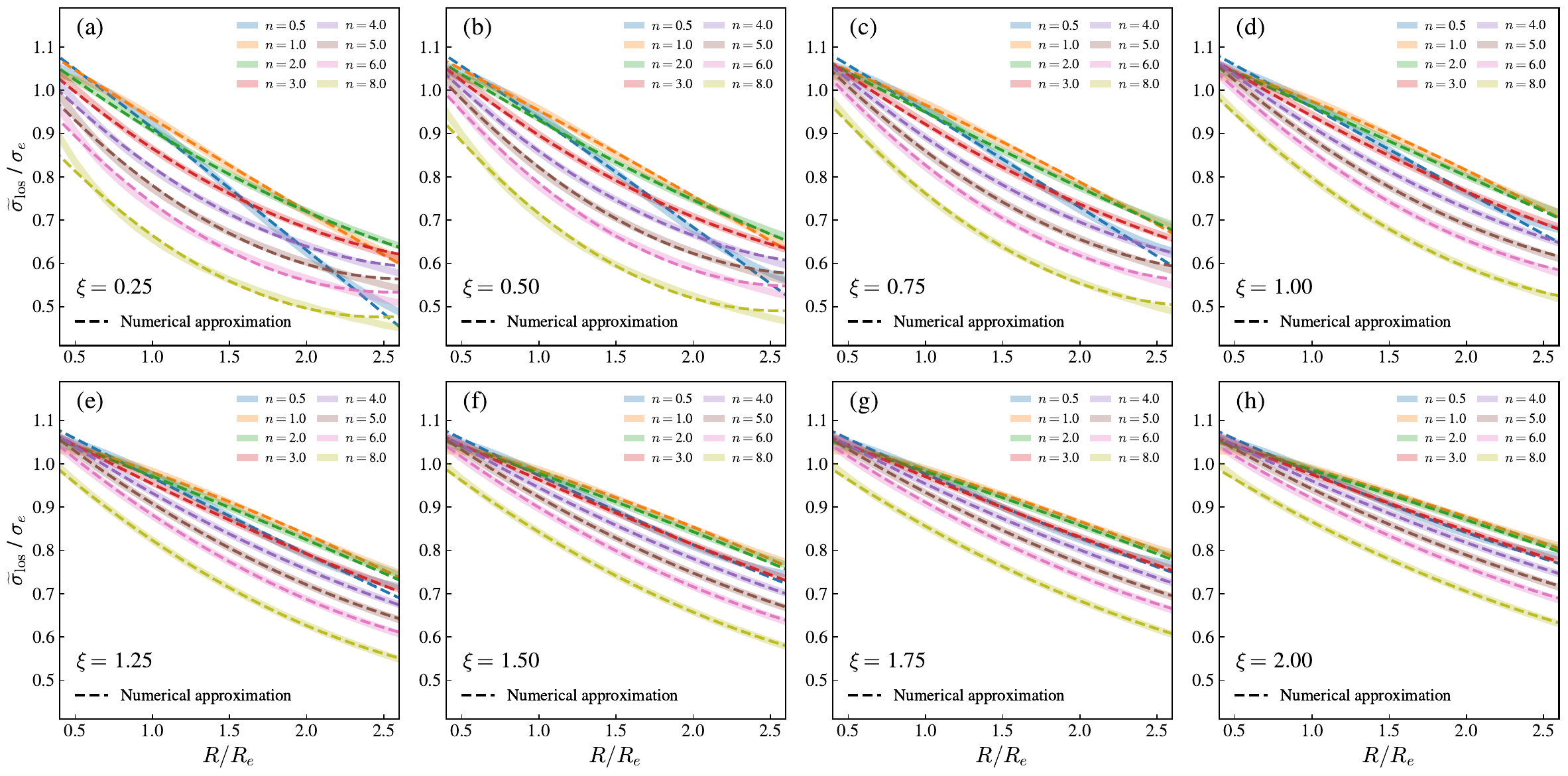}
\caption{\label{fig:ap_corr_conv} Annular aperture correction factor as a function aperture radial location. Panels (a) to (h) show particular cases of PSF FWHM values normalized to S\'ersic model $R_e$. The numerical approximation for annular aperture correction factor corresponds to that of Equation~\ref{eq:ap_corr}, adopting the input coefficients presented in Table~\ref{tab:ap_corr_constants}.}
\end{figure*}

Figure~\ref{fig:ap_conv_effect} shows the effect of PSF convolution on the recovery of $\sigma_e$. The overall effect is a decrease of the annular aperture correction factor that must be applied when the PSF FWHM increases relative to $R_e$, as expected due to the spreading of the central emission. This conclusion applies for values of S\'ersic index beyond the $n=1$ and 4 cases presented here. Figure~\ref{fig:ap_corr_conv} presents the annular aperture correction factors as function of radius, normalized to $R_e$, for different S\'ersic index $n$, and various ratios of $\xi \equiv$\,PSF FWHM/$R_e$. The annular aperture correction factor becomes less sensitive to the bulge S\'ersic index for large $\xi$ values, indicating that determining the underlying surface brightness profile is less critical.

Given that computing $F_{\rm ap}$ is considerably time intensive, we develop a cheaper numerical parameterization that can be used once the bulge profile and observation PSF have been characterized. The numerical parameterization is detailed in Equations~\ref{eq:ap_corr} and \ref{eq:ap_corr_coeff}, with input coefficients given in Table~\ref{tab:ap_corr_constants}. These coefficients were derived by fitting $F_{\rm ap}$ in a radius range from 0.5 to $2.5\,R_e$ in steps of $0.1\,R_e$, $n = 0.5-8$ varying in bins 0.5 wide, and $\xi = 0-2.0$ sampled in steps of 0.25. This numerical recipe is accurate up to $2\%$, sufficient given the larger uncertainties expected from modeling AGN host galaxy bulges. We note that, with the exception of the case of no PSF blurring ($\xi = 0$), our numerical recipe in only valid when the observation PSF is well-described by a Moffat profile. However, the procedure outlined here can be adopted for different representations of the PSF.

\end{appendix}

\end{document}